\documentclass[manuscript]{aastex}

\usepackage{natbib}
\usepackage{epsfig}

\shorttitle{Seven binaries discovered through the BEER method}
\shortauthors{Faigler et al.}

\begin{document}

\title{Seven new binaries discovered in the {\it Kepler} light curves through the BEER method confirmed by radial-velocity observations}

%% Use \author, \affil, and the \and command to format
%% author and affiliation information.
%% Note that \email has replaced the old \authoremail command
%% from AASTeX v4.0. You can use \email to mark an email address
%% anywhere in the paper, not just in the front matter.
%% As in the title, use \\ to force line breaks.

\author{S. Faigler and T. Mazeh}
\affil{ School of Physics and Astronomy, Raymond and Beverly Sackler Faculty of Exact Sciences, Tel Aviv University, Tel Aviv  69978, Israel}

\and
\author{S. N. Quinn and D. W. Latham}
\affil{Harvard-Smithsonian Center for Astrophysics,60 Garden St., Cambridge, MA 02138}

\and
\author{L. Tal-Or}
\affil{ School of Physics and Astronomy, Raymond and Beverly Sackler Faculty of Exact Sciences, Tel Aviv University, Tel Aviv  69978, Israel}

\begin{abstract}
We present seven newly discovered {\it non-eclipsing} short-period binary systems with low-mass companions, identified by the recently introduced BEER algorithm, applied to the publicly available 138-day photometric light curves obtained by the {\it Kepler} mission.  
The detection is based on the beaming effect (sometimes called Doppler boosting), which increases (decreases) the brightness of any light source approaching (receding from) the observer, 
enabling a prediction of the stellar Doppler radial-velocity modulation from its precise photometry.  The BEER algorithm identifies the BEaming periodic modulation, with a combination of the well known Ellipsoidal and Reflection/heating periodic effects, induced by short-period companions. The seven detections were confirmed by spectroscopic radial-velocity follow-up observations, indicating minimum secondary masses in the range of $0.07$--$0.4$ $M_{\odot}$.  
The discovered binaries establish for the first time the feasibility of the BEER algorithm as a new detection method for short-period non-eclipsing binaries, with the potential to detect in the near future non-transiting brown dwarfs secondaries, or even massive planets. 
\end{abstract}

%% Keywords should appear after the \end{abstract} command. The uncommented
%% example has been keyed in ApJ style. See the instructions to authors
%% for the journal to which you are submitting your paper to determine
%% what keyword punctuation is appropriate.


\section{Introduction}

In a recent paper \citet{faigler11} presented a new way to discover short-period non-eclipsing binaries with low-mass companions by using highly precise photometric light curves obtained by space missions, like CoRoT and {\it Kepler} \citep{rouan98, baglin06, borucki10}. The algorithm, BEER, based on an idea suggested by \citet{loeb03} and \citet{zucker07}, searches for the beaming effect, sometimes called Doppler boosting, induced by stellar radial motion.
This effect causes an increase (decrease) of the brightness of any light source approaching (receding from) the observer \citep{rybicki79}, on the order of $4v_r/c$, where $v_r$ is the radial velocity of the source, and $c$ is the velocity of light. Therefore, periodic modulation of the stellar velocity due to a companion in a binary orbit will produce a corresponding periodic beaming modulation of the stellar photometry.  

For short-period binaries the beaming effect is extremely small, on the order of 100--300  ppm (parts per million). Therefore the effect has become relevant only recently, when CoRoT and {\it Kepler} --- the two presently operating satellites that search for transiting exoplanets,
started producing hundreds of thousands of uninterrupted light curves with high precision \citep{auvergne09, koch10}.

As predicted, several studies detected the beaming effect in eclipsing binaries and transiting planets, for which the orbital period was well established from the space-obtained light curves \citep{vankerkwijk10,rowe10,carter10,mazeh10,bloemen11,kipping11}. Yet, space mission data can yield much more.  Evidence of the binarity of a stellar system can be found from detecting the beaming effect without any eclipse or transit \citep{loeb03, zucker07}. However, the beaming modulation by itself might not be enough to render a star a good binary candidate, as periodic modulations could be
produced by other effects, stellar variability in particular \citep{aigrain04}.

The BEER detection algorithm \citep{faigler11}, therefore, searches for stars that show in their light curves a combination of the BEaming effect with two other effects induced by the presumed companion --- the Ellipsoidal and the Reflection modulation. The ellipsoidal variation \citep{morris85} is due to the tidal distortion of each component by the gravity of its companion (see a review by \citet{mazeh08}), while the reflection/heating variation (referred to herein as the reflection modulation) is induced by the luminosity of each component that falls only on the close side of its companion \citep{vaz85,wilson90,maxted02,harrison03,for10,reed10}.
Detecting the beaming effect together with the ellipsoidal and reflection modulations, with the expected relative amplitudes and {\it phases} in particular, can suggest the presence of a small non-transiting companion. 

Just as transit searches, the candidates found by the BEER algorithm
have to be followed by radial-velocity (RV) observations, in order to confirm the existence
of the low-mass companion, and to reject other possible interpretations of the photometric modulation. 

This paper presents the discovery of the first seven new binaries with low-mass secondaries, in the range of $0.07$--$0.4\, M_{\odot}$, detected by using the BEER algorithm, and confirmed by RV spectroscopic follow-up measurements. Section~2 presents the photometric analysis of the {\it Kepler} light curves, Section~3 provides the details and results of the RV observations, Section~4 summarizes and compares the results of the photometric analysis and the RV measurements, and Section~5 discusses the implications of, and conclusions from, the findings of this paper. 

%====================
\section{Photometric analysis}
%====================

We used the publicly available {\it Kepler} raw light curves of the Q0, Q1 and Q2 quarters, spanning $138$ days. To avoid systematic variations, we ignored all data points within $1$ day after the beginning of Q2, and all data points within $1$ day before, to $3$ days after, each of the two safe mode events in Q2. We also corrected two systematic jumps at {\it Kepler} time ($JD-2454833$) of $200.32$ and $246.19$ days. We then applied the BEER algorithm to
$14{,}685$ stars brighter than $13$th mag, with {\it Kepler} Input Catalog \citep{brown11} radius smaller than $3R_{\odot}$, calculating the BEER periodogram \citep{faigler11} with period range of $0.5$--$20$ days for each star. 
Next, we identified the periodograms whose highest peak was at least 3 time higher than the next highest one. For these stars we used the peak period to estimate the system secondary mass and radius, assuming the periodicity is induced by a secondary star.  We then selected 25 candidates with secondary mass smaller then $0.5M_{\odot}$ and implied albedo smaller then $0.4$, 
suggesting a significant probability for a low-mass companion. These candidates were then followed by RV observations, which we describe in detail in the next section. In a forthcoming paper we will report on the false alarm cases, and analyze the false alarm frequency of our candidates. Here we report on the first seven clear detections. 

Table~1 lists for each of the seven stars its coordinates, the stellar properties estimates from the Kepler Input Catalog \citep{brown11}, the photometric periods and amplitudes of the three effects found by the BEER algorithm, and the r.m.s. of the data before and after subtraction of the BEER model.

 We order the stars according to the detected RV amplitude, presented in the next section. Figure~1 presents the `cleaned' \citep{mazeh10,faigler11} photometric data of the seven detections,  Figure~2 presents  the BEER periodograms for the detections, and Figure~3 shows the light curves folded with the detected period.
In fact, the quality of the {\it Kepler} data is so high that the periodic modulation can be seen directly from the cleaned data, plotted in Figure~1, even without consulting the BEER periodogram. 

It is interesting to compare the shape of the BEER modulation of the seven candidates, presented in Fig~3. In six of them the two peaks, at phase of 0.25 and 0.75, are similar, although the latter is somewhat smaller, due to the beaming effect \citep{faigler11}. In one case, K08016222, the second peak completely disappeared, because in this case the beaming amplitude is more than three times higher than that of the ellipsoidal, while for the rest of the candidates, the ellipsoidal amplitude is significantly higher then the beaming amplitude. This is a clear result of the long orbital period and small stellar radius of this system, relative to the other systems, since the ellipsoidal amplitude to beaming amplitude ratio is proportional to $R_*^{3}/P_{orb}^{5/3}$  \citep{faigler11,zucker07}.

%  Table 1
%-------------
%
\begin{deluxetable}{lrrrrrrr}
\tabletypesize{\scriptsize}
%\begin{table}
%\rotate
\tablecaption{ Coordinates, magnitudes, stellar properties, and photometric analysis results of the seven candidates}

\tablewidth{0pt}
\tablehead{
& \colhead{K10848064}  & \colhead{K08016222} &  \colhead{K09512641} & \colhead{K07254760} & \colhead{K05263749} & \colhead{K04577324} & \colhead{K06370196}
}

\startdata
RA  & 19:01:21.24 & 19:06:48.03	& 18:58:39.91 & 18:42:28.78 & 19:12:59.00 & 19:42:35.91 & 19:35:00.36 \\
DEC &48:16:32.90  & 43:48:32.90 & 46:08:52.80 & 42:49:31.90 & 40:26:42.30 & 39:38:00.80 & 41:47:59.60 \\
\tableline
${K_{p}}^{a}$ [mag] &12.13  & 11.65 & 11.66 & 12.04 & 11.53 & 11.98 & 11.97 \\
$R$$^a$ [$ R_{\odot}$] & $1.5$   & $1.3$  & $1.7$  & $1.5$  & $1.9$  & $1.3$  & $2.1$                  \\
$M$$^b$ [$ M_{\odot}$] & $1.2$   & $1.1$  & $1.2$  & $1.2$  & $1.3$  & $1.2$  & $1.3$                  \\

\tableline
Photometry results:\\

Period [days]    & $3.49 \pm 0.01$ & $5.60 \pm 0.02$ & $4.65 \pm 0.02$ & $2.66 \pm 0.01$ & $3.73 \pm 0.01$ & $2.33 \pm 0.01$ & $4.23 \pm 0.01$ \\ 

Ellipsoidal [ppm]  & $201 \pm 3$ & $30 \pm 2$ & $172 \pm 10$ & $845 \pm 7$ & $1222 \pm 5$ & $1489 \pm 4$ & $1210 \pm 10$       \\

Beaming [ppm]   & $118 \pm 3$ & $97 \pm 2$ & $185 \pm 5$ & $356 \pm 6$ & $358 \pm 5$ & $436 \pm 4$ & $382\pm 7$           \\

Reflection [ppm]  & $0 \pm 3$ & $6 \pm 2$ & $36 \pm 4$ & $150 \pm 6$ & $158 \pm 5$ & $245 \pm 4$ & $174 \pm 7$           \\

Cleaned data r.m.s. [ppm]  & $204$ & $106$ & $227$ & $807$ & $1184$ & $1408$ & $1141$           \\
%$\sigma$
Residuals r.m.s. [ppm]       & $128$ & $68$ & $113$ & $268$ & $200$ & $168$ & $328$           \\

\enddata
\tablenotetext{\space}{ $^a$  from Kepler Input Catalog}
\tablenotetext{\space}{ $^b$  calculated from Kepler Input Catalog $\log g$ and $R$  }
\end{deluxetable}

%\label{table_stars}

%---------------------------
% Figure 1%
%---------------------------
\begin{figure*} 
\centering
\resizebox{17cm}{15cm}
{
 \includegraphics{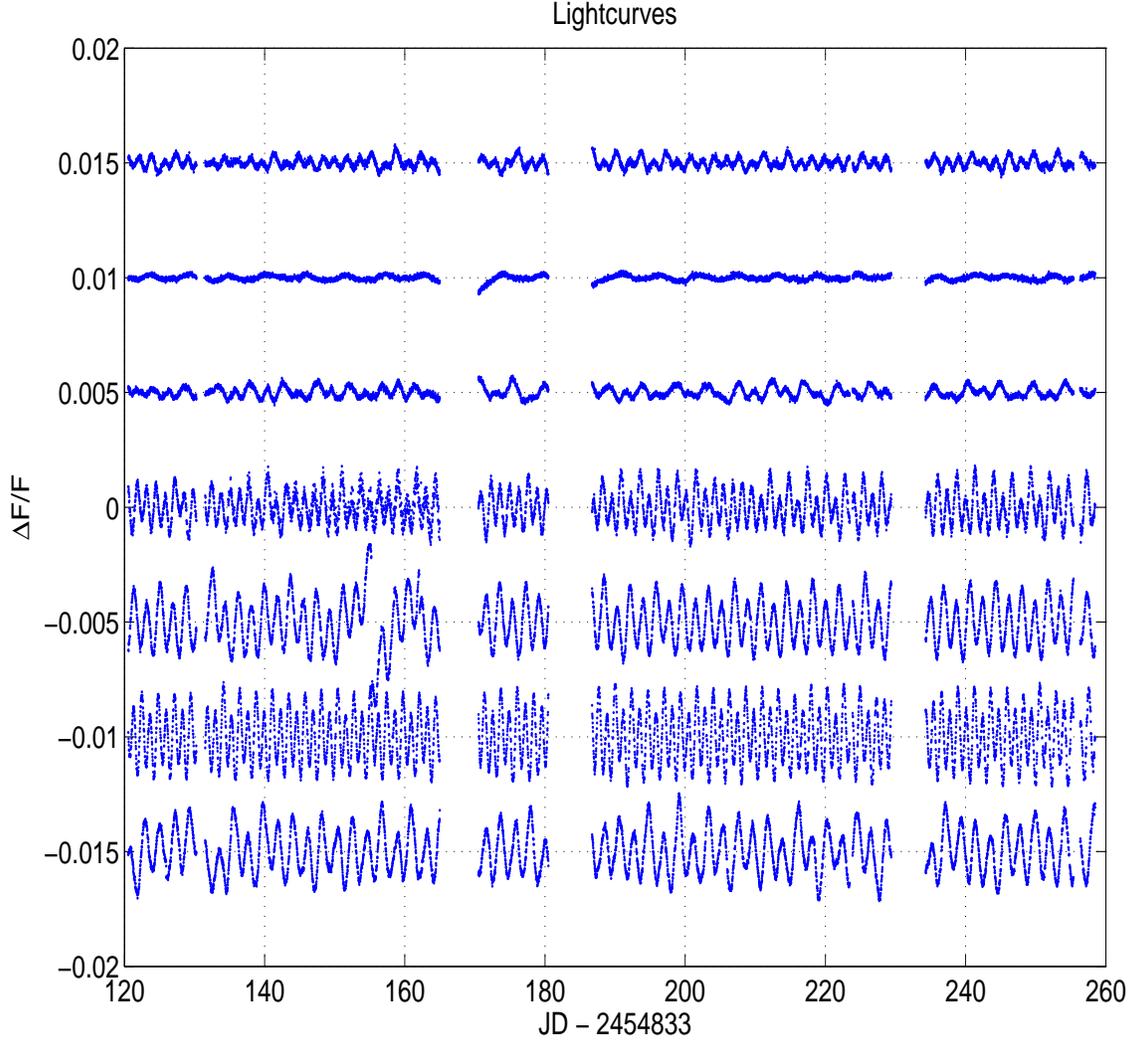}  
}

\caption{
The light curves of the seven detections, after outlier removal and long-term detrending. 
Top to bottom:  K10848064, K08016222 , K09512641, K07254760, K05263749, K04577324, K06370196.
For clarity, each light curve was shifted by $5000$ ppm relative to the previous one.
The periodic modulation can be seen in all seven light curves.
The light curves show several discontinuities: end of Q0 at day 131, end of Q1 around day 166, Q2 first safe mode event at day 183, and Q2 second safe mode event at day 232.
In addition there is a single discontinuity at day 155 of the K05263749 light curve.
}
\end{figure*}
%--------------------------------------------------------------------------------

%---------------------------
% Figure 2%
%---------------------------
\begin{figure*} 
\centering
\resizebox{16cm}{5.33cm}
{
\includegraphics{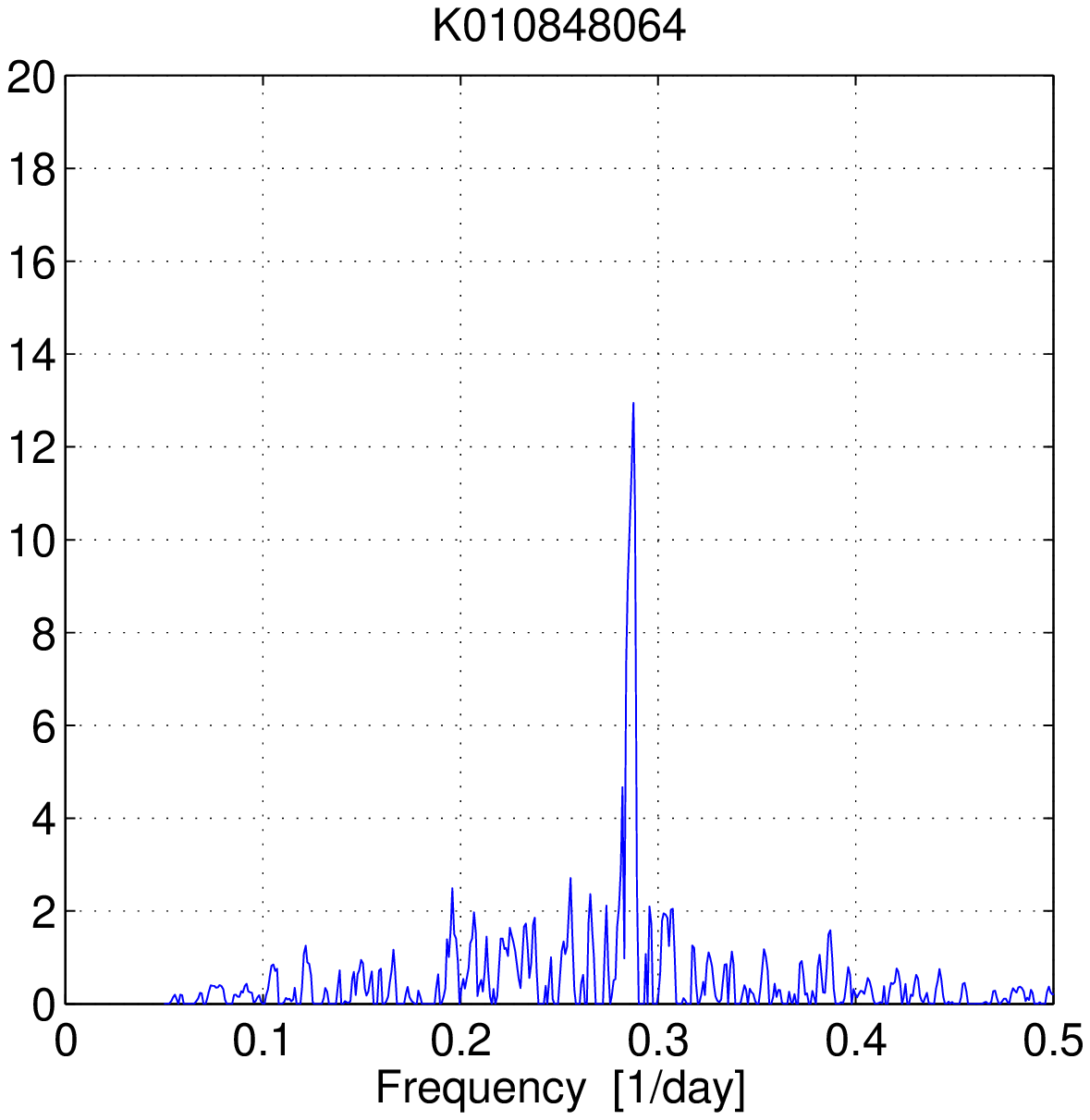}  
 \includegraphics{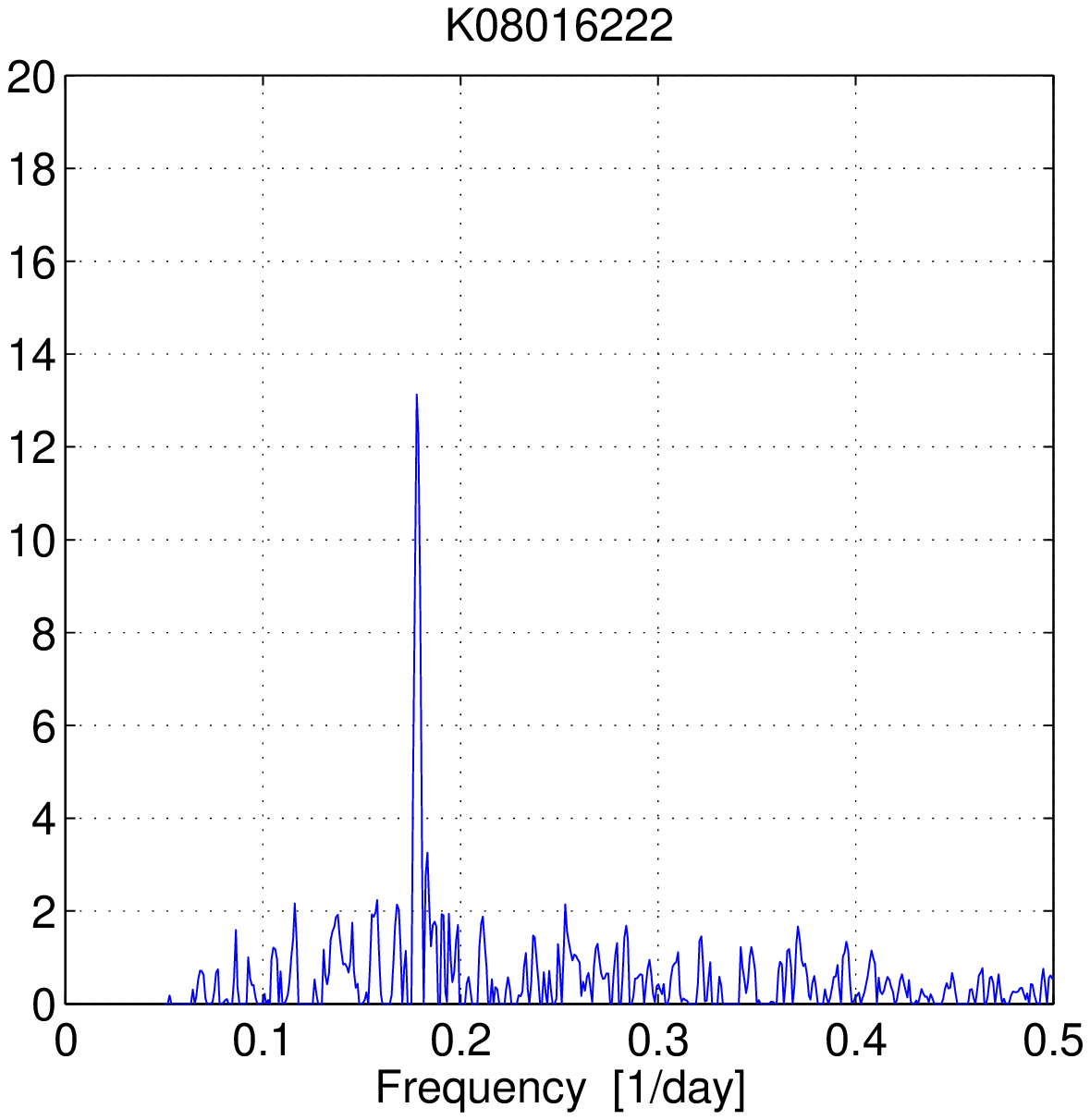}
 \includegraphics{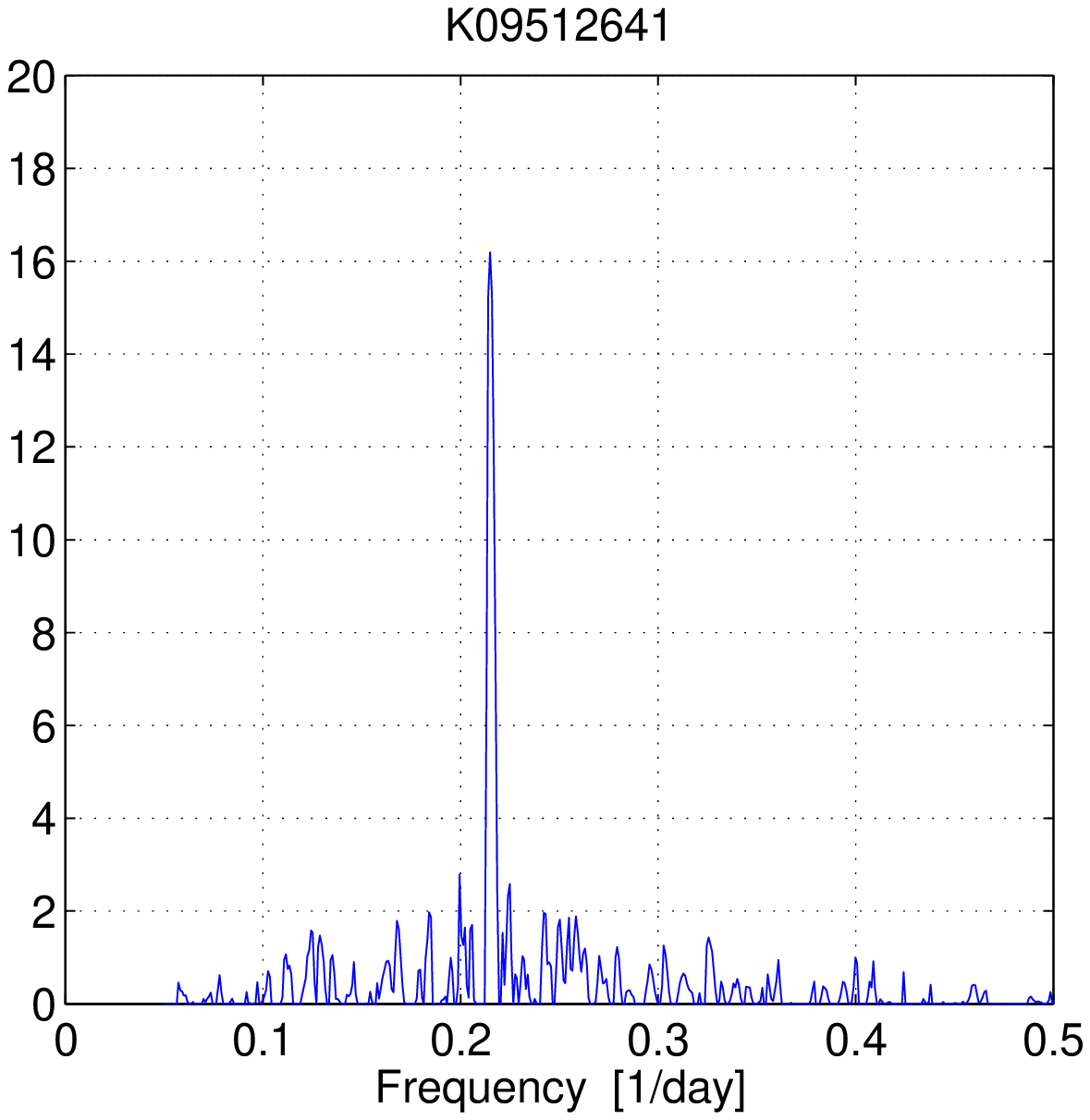}
}

\centering
\resizebox{16cm}{5.33cm}
{
 \includegraphics{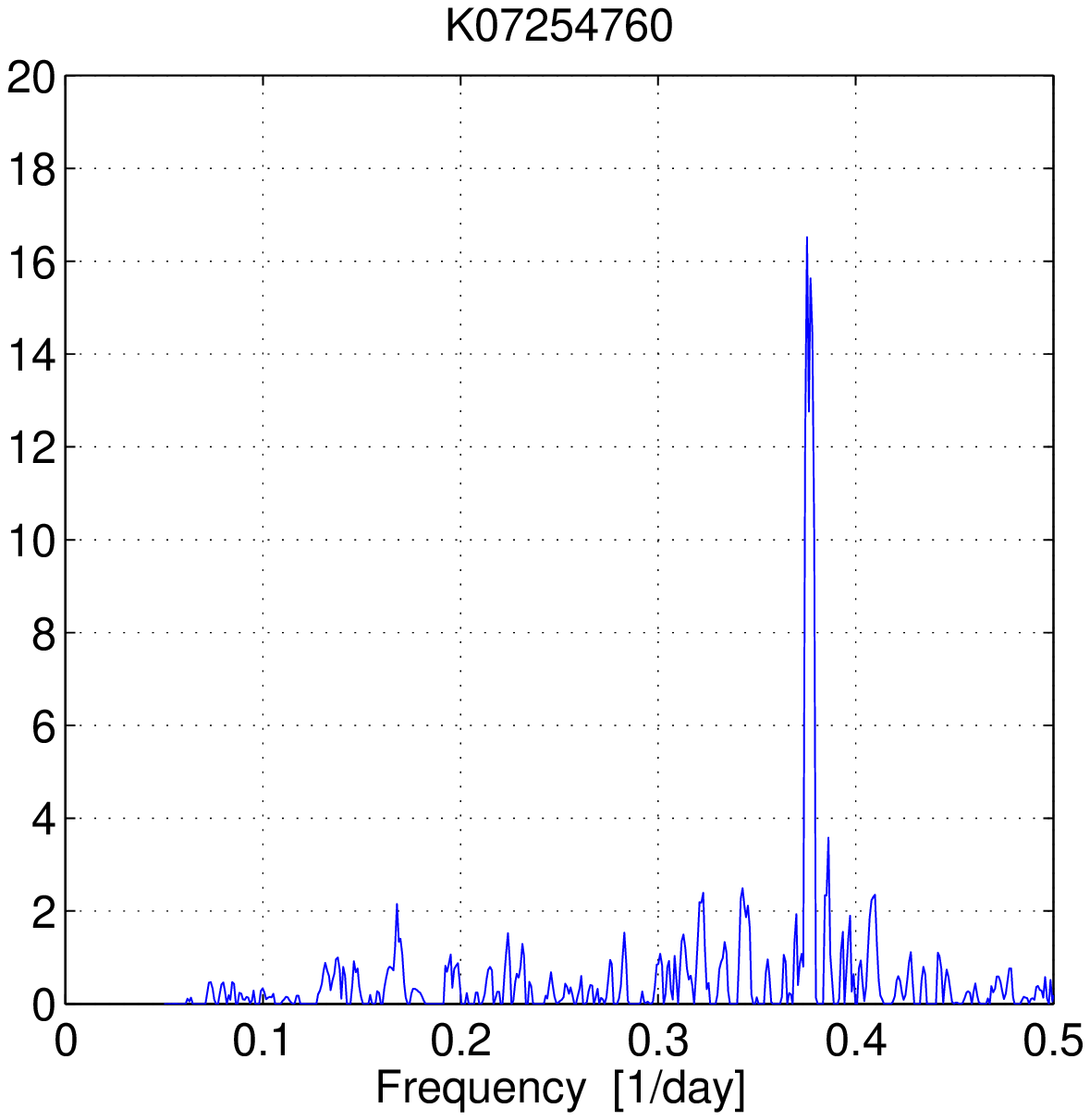}  
 \includegraphics{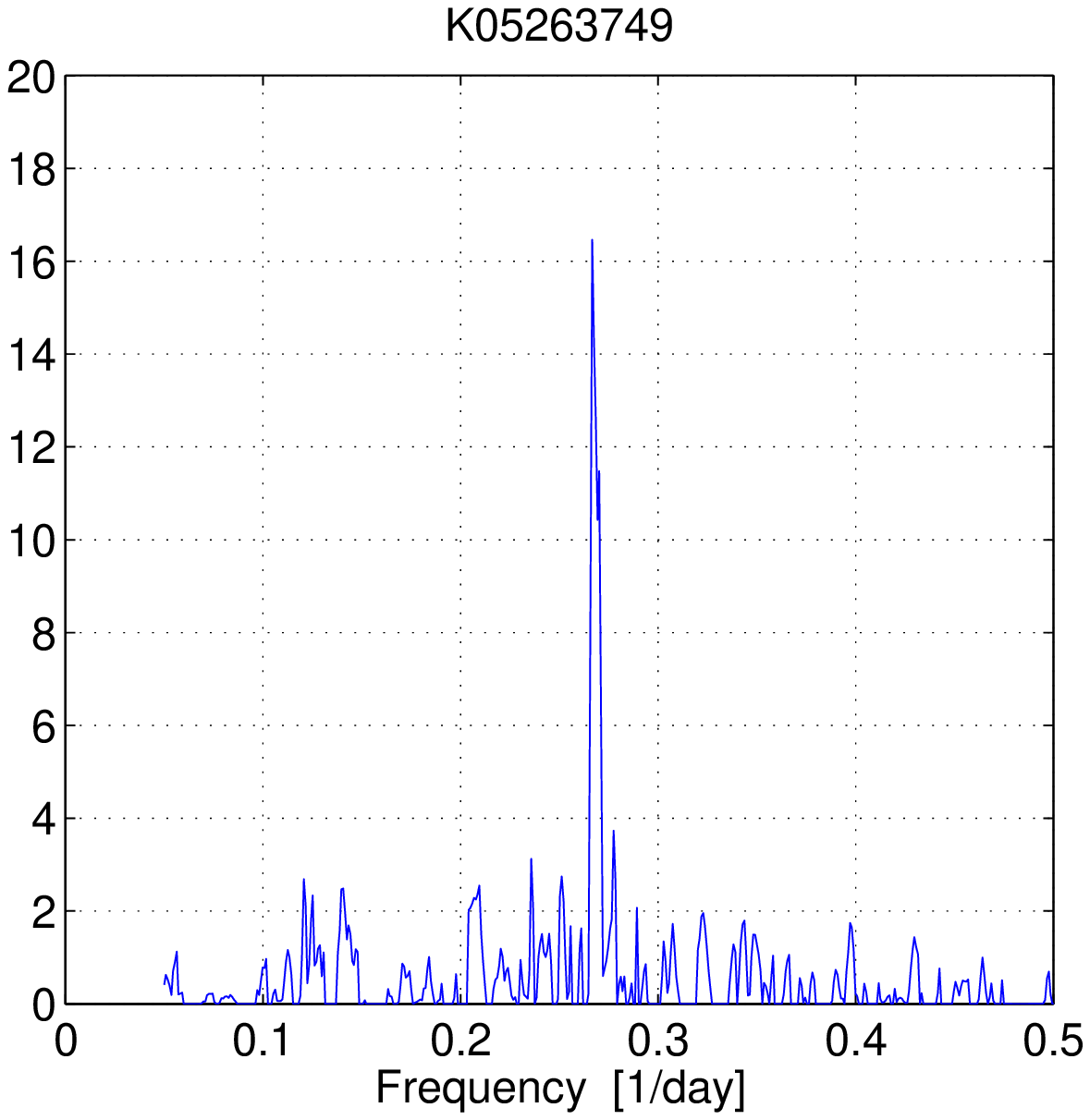}
 \includegraphics{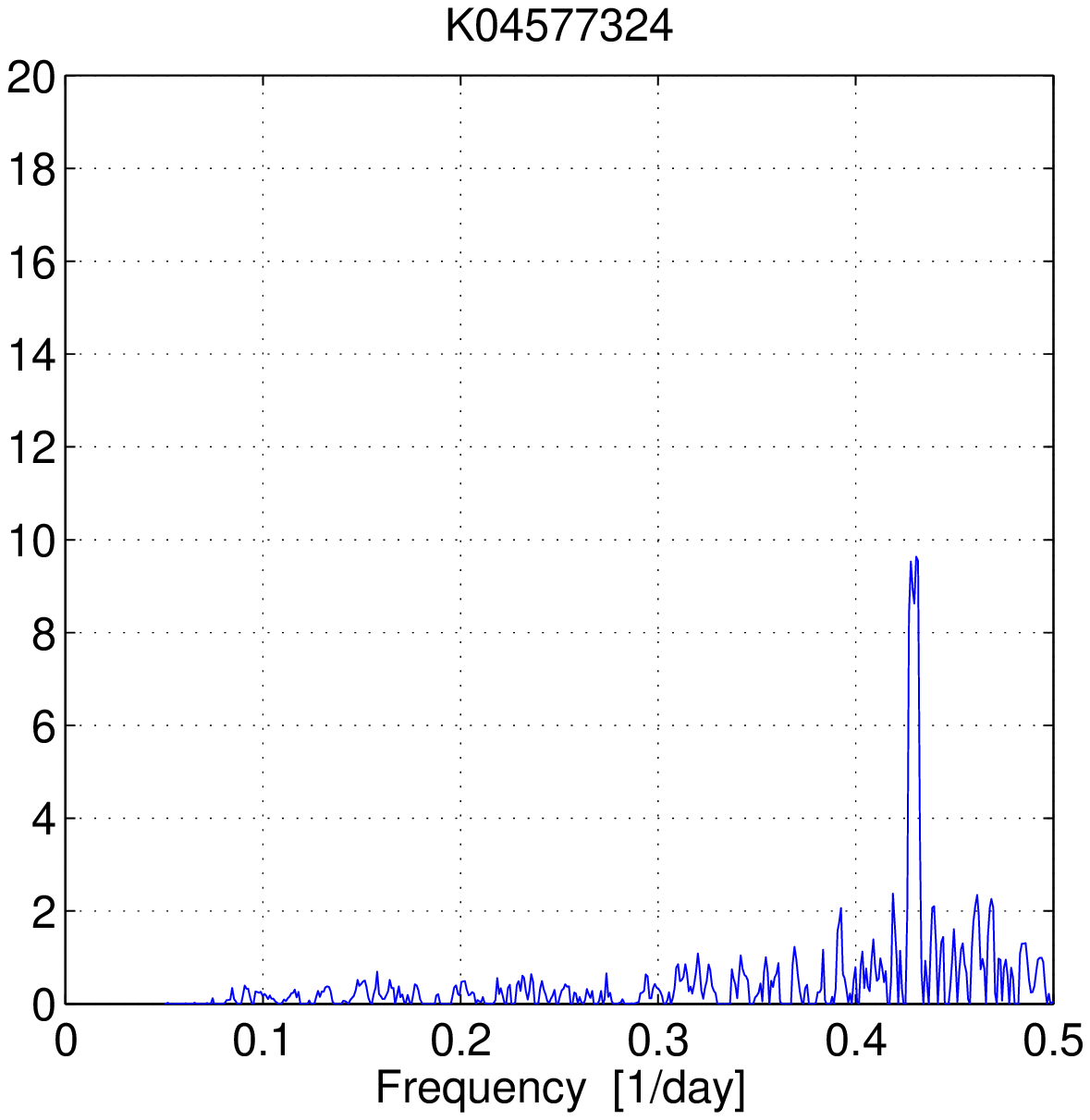}
}

%{\oddsidemargin 3cm
\centering
\resizebox{5.33cm}{5.33cm}
{
 \includegraphics{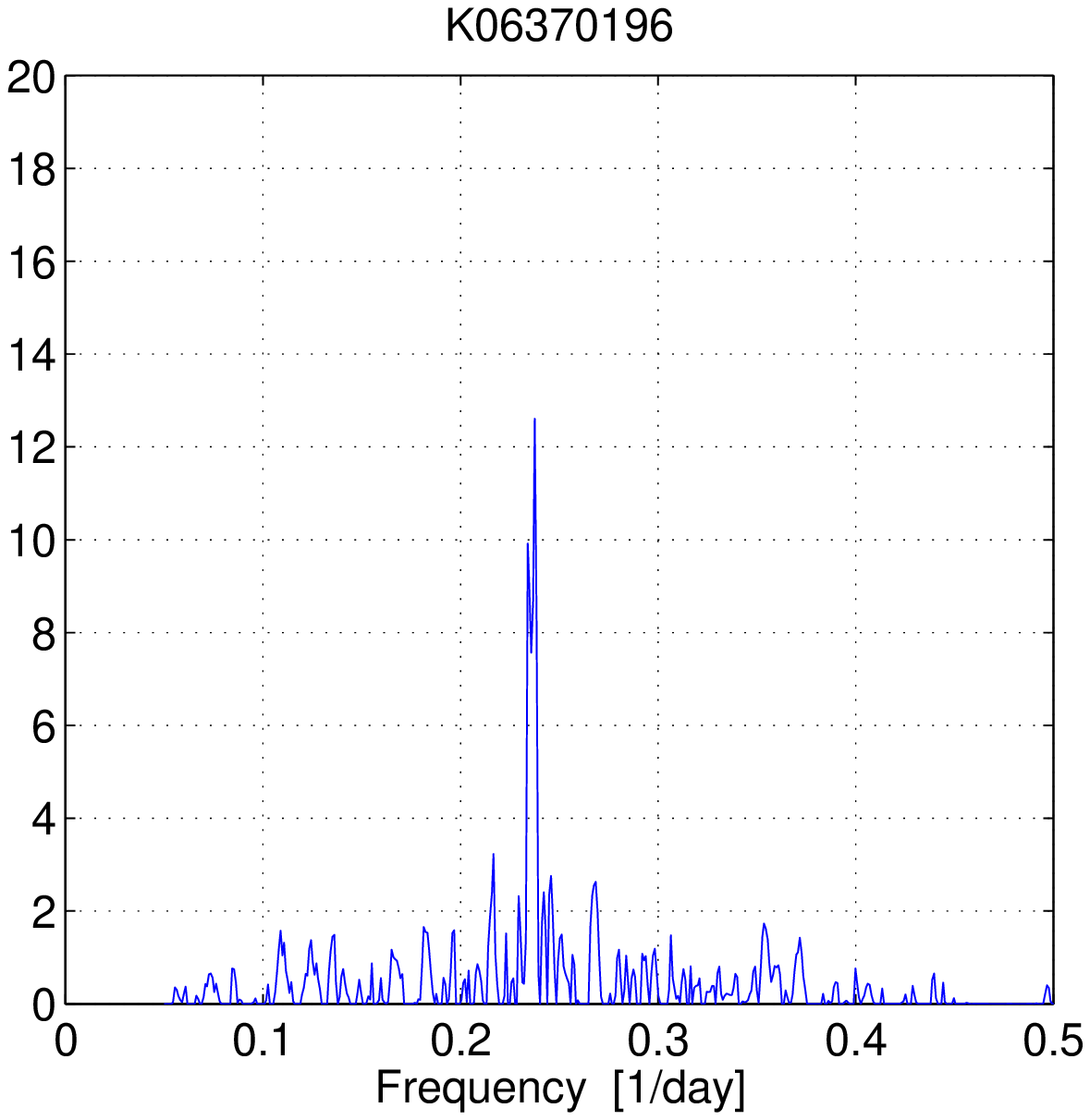}  
 }
\caption{
The BEER periodograms of the seven candidates. The peak frequency
corresponds to the suspected orbital period. 
%{\bf The periodograms were calculated for periods range of $0.5$--$20$ days, but are plotted here, for clarity purposes, only for periods range of $2$--$20$ days. This is because no significant peak exists for periods smaller then $2$ days in any of the periodograms.}
 The periodograms were calculated for the period range of $0.5$--$20$ days. For clarity, only the period range of $2$--$20$ days is plotted, since no significant peak was found for periods smaller than $2$ days in any of the periodograms.
The periodograms are normalized so that the r.m.s. of the $100$ noise points on two sides of the peak ($50$ on each side) is set to one.
}
\end{figure*}
%--------------------------------------------------------------------------------

%---------------------------
% Figure 3%
%---------------------------
\begin{figure*} 

\centering
\resizebox{16cm}{5.33cm}
{
%\subfloat{\label{fig:1}\includegraphics[width=0.3\textwidth]{K8016222_fold.eps}}  
%\includegraphics{seven_fold.eps}  
%\begin{center}
 \includegraphics{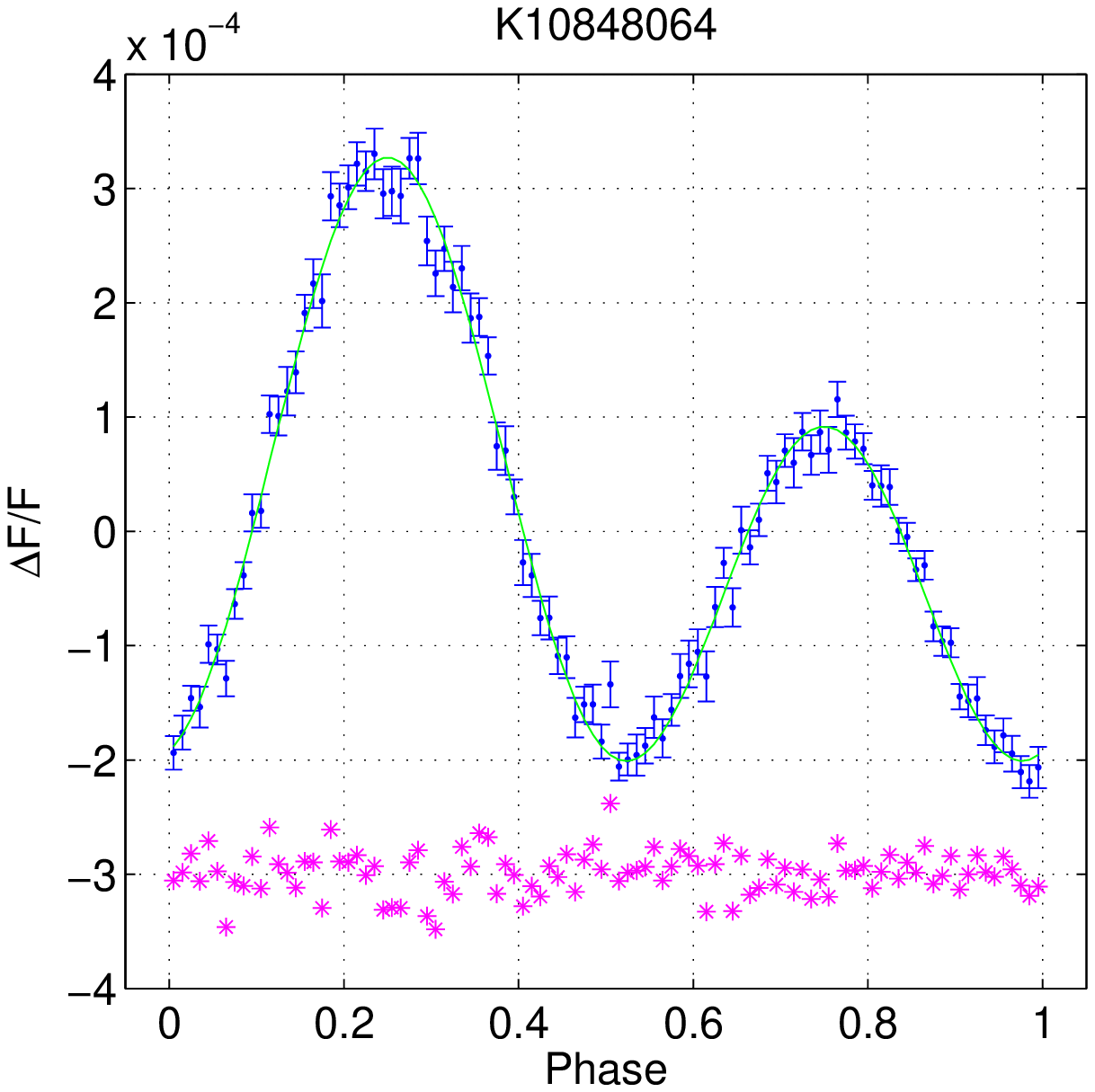}  
 \includegraphics{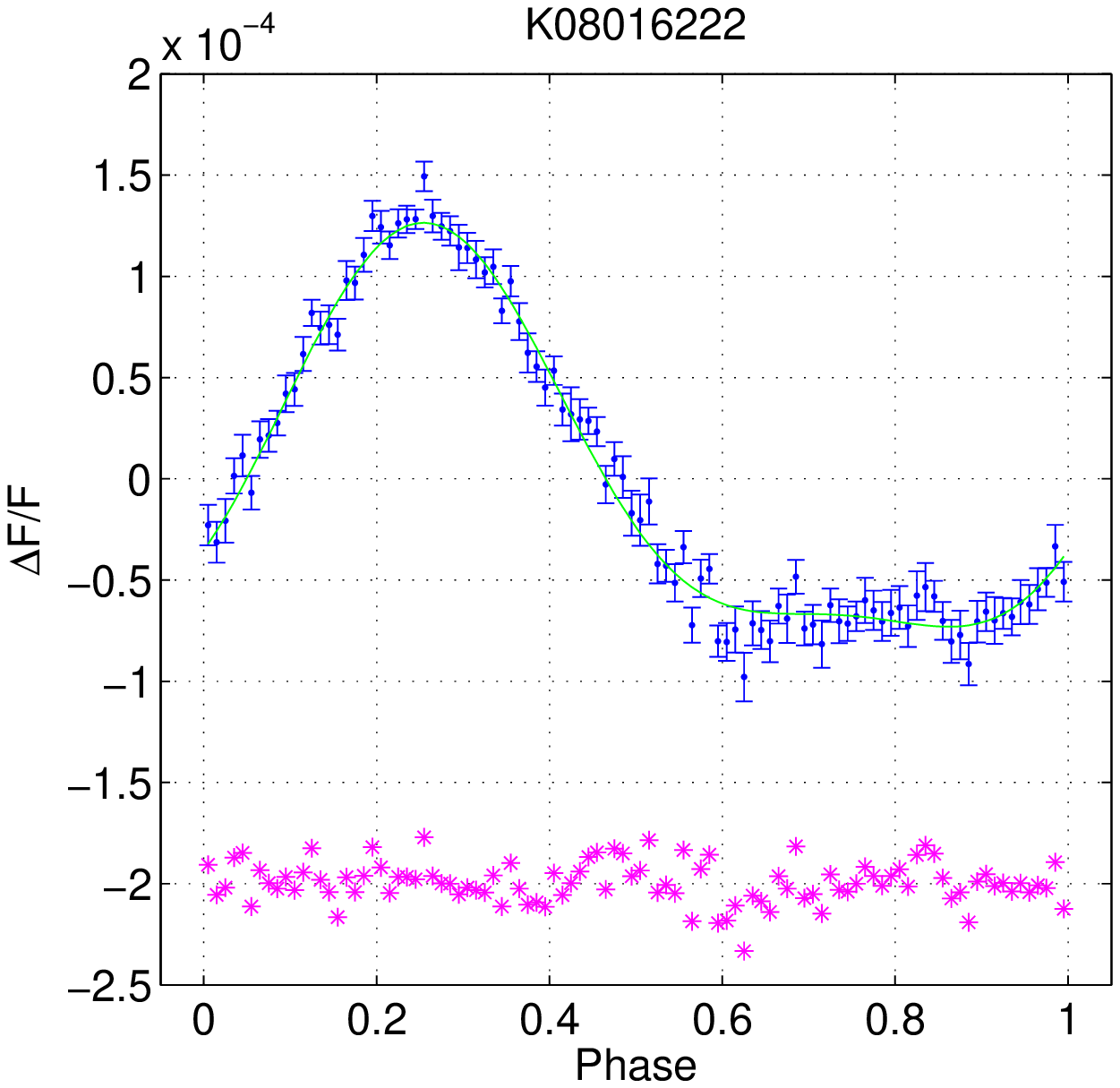}
 \includegraphics{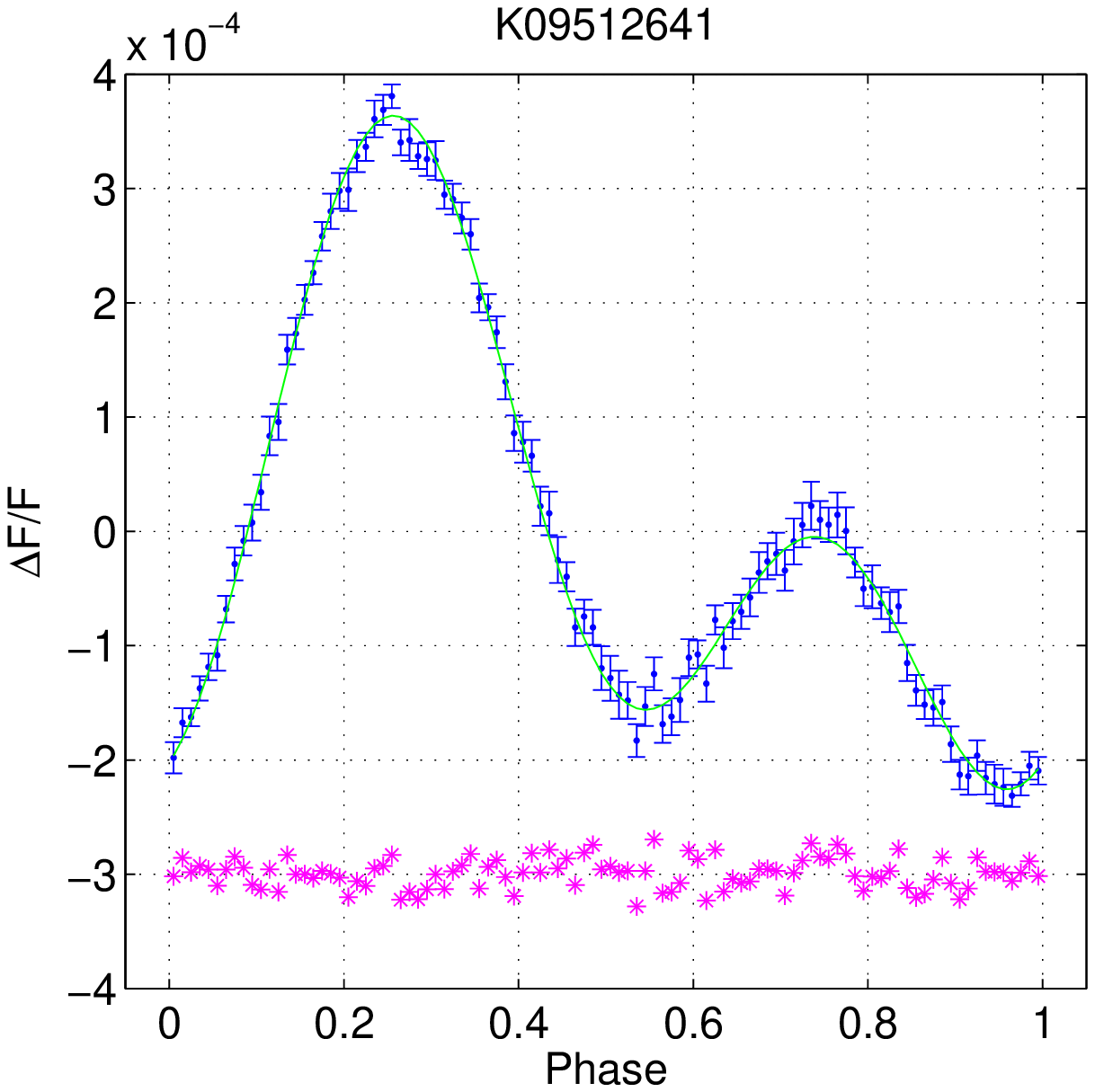}
}

\centering
\resizebox{16cm}{5.33cm}
{
 \includegraphics{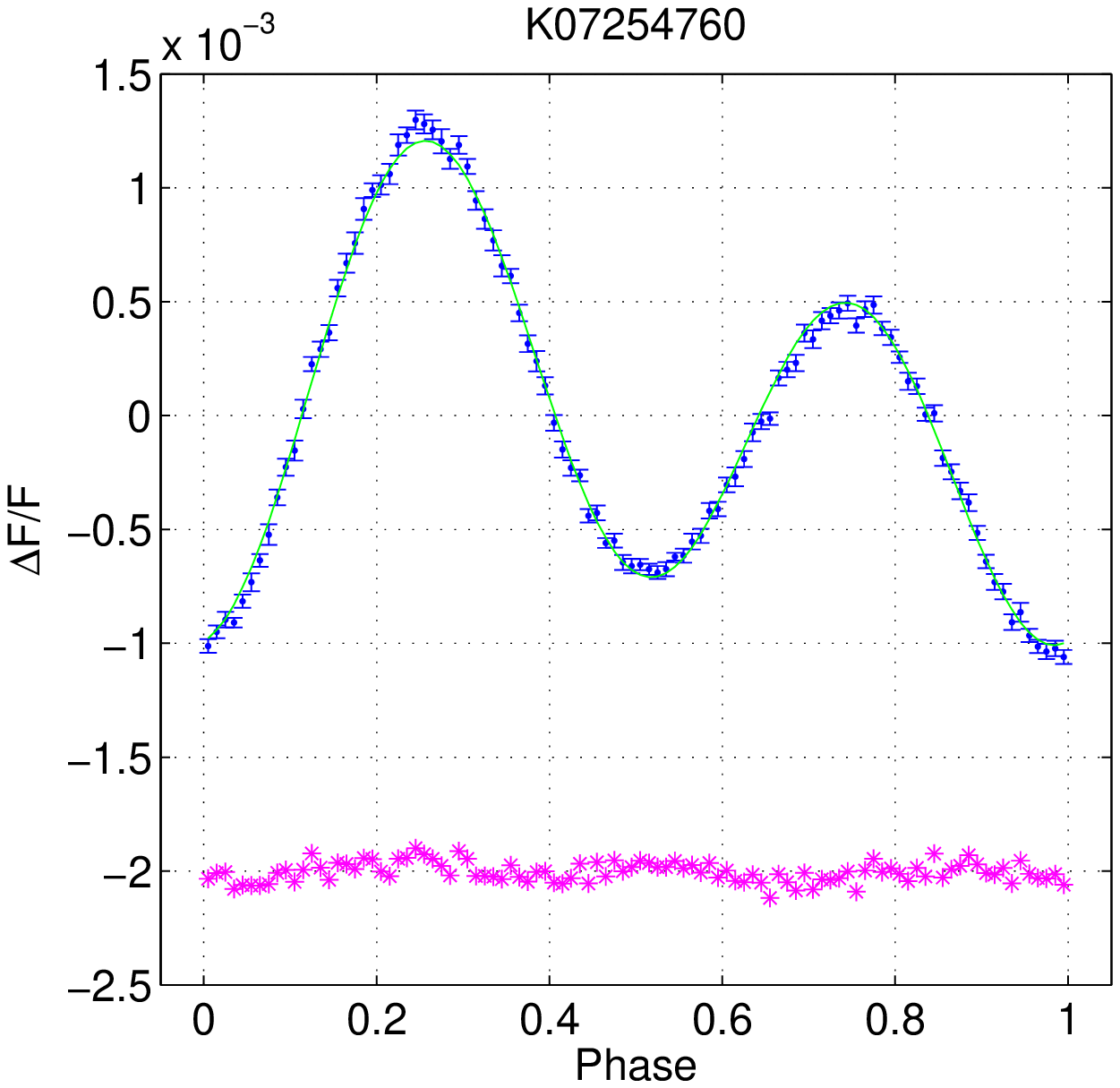}  
 \includegraphics{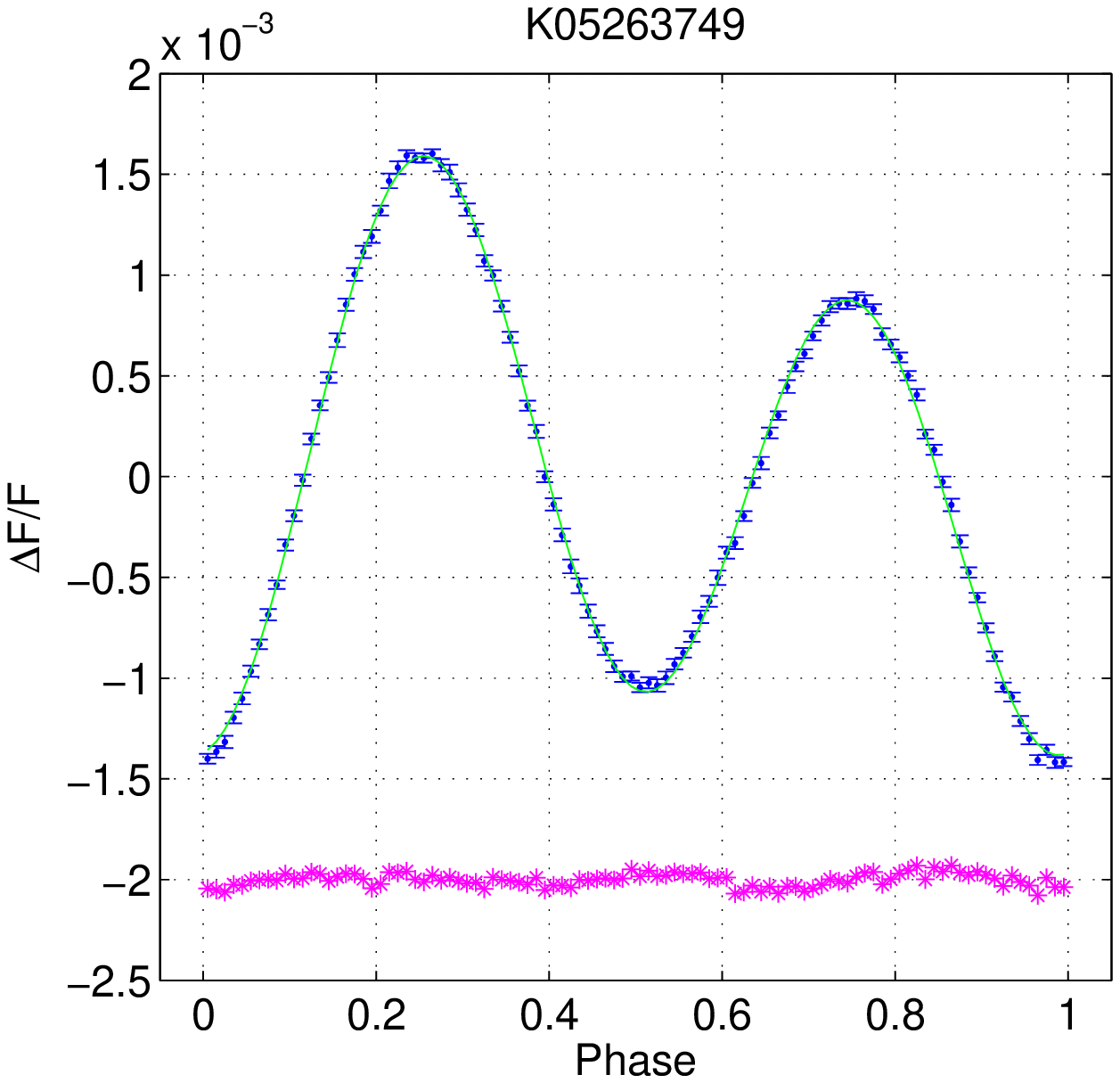}
 \includegraphics{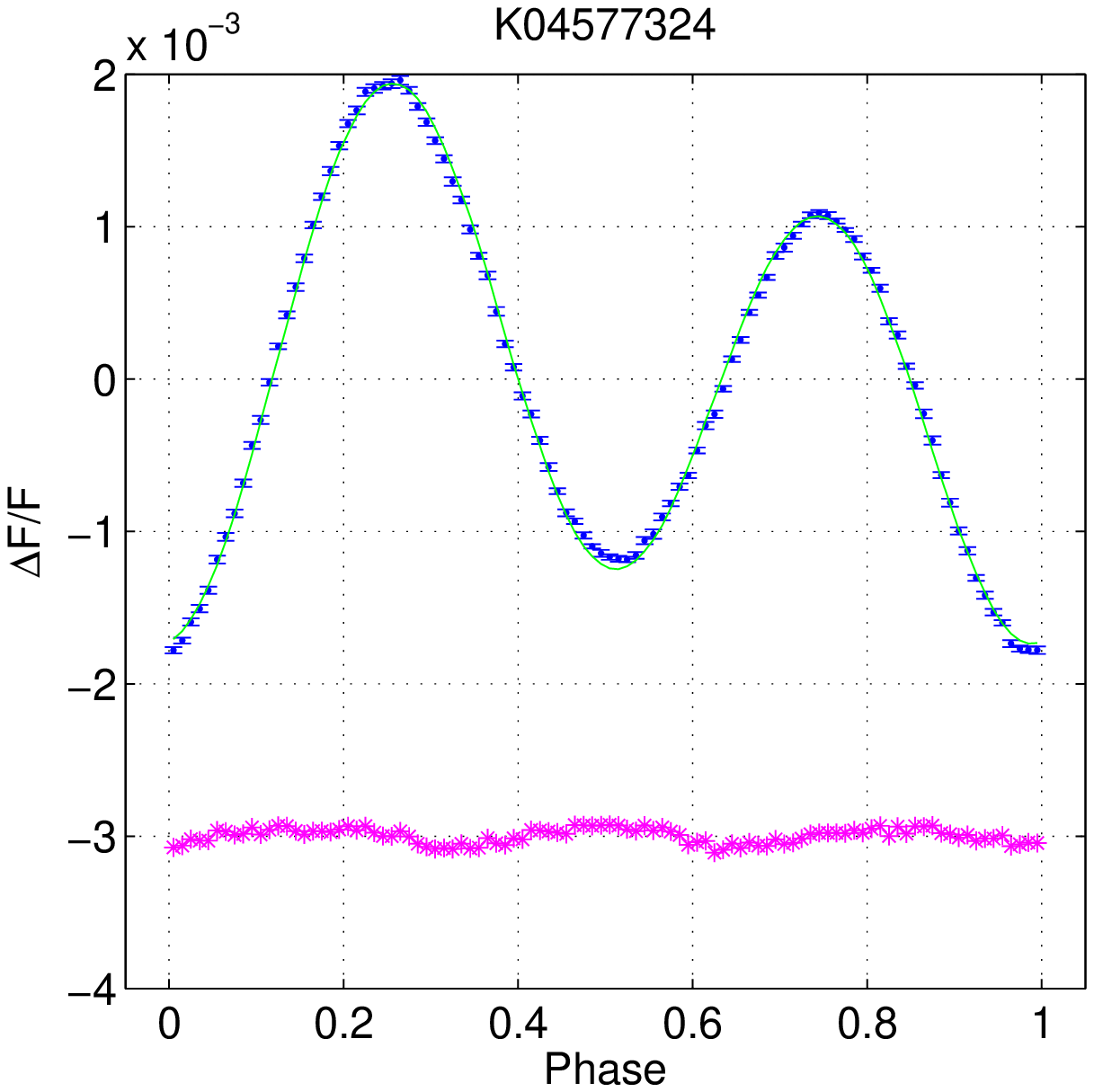}
}

\centering
\resizebox{5.33cm}{5.33cm}
{
 \includegraphics{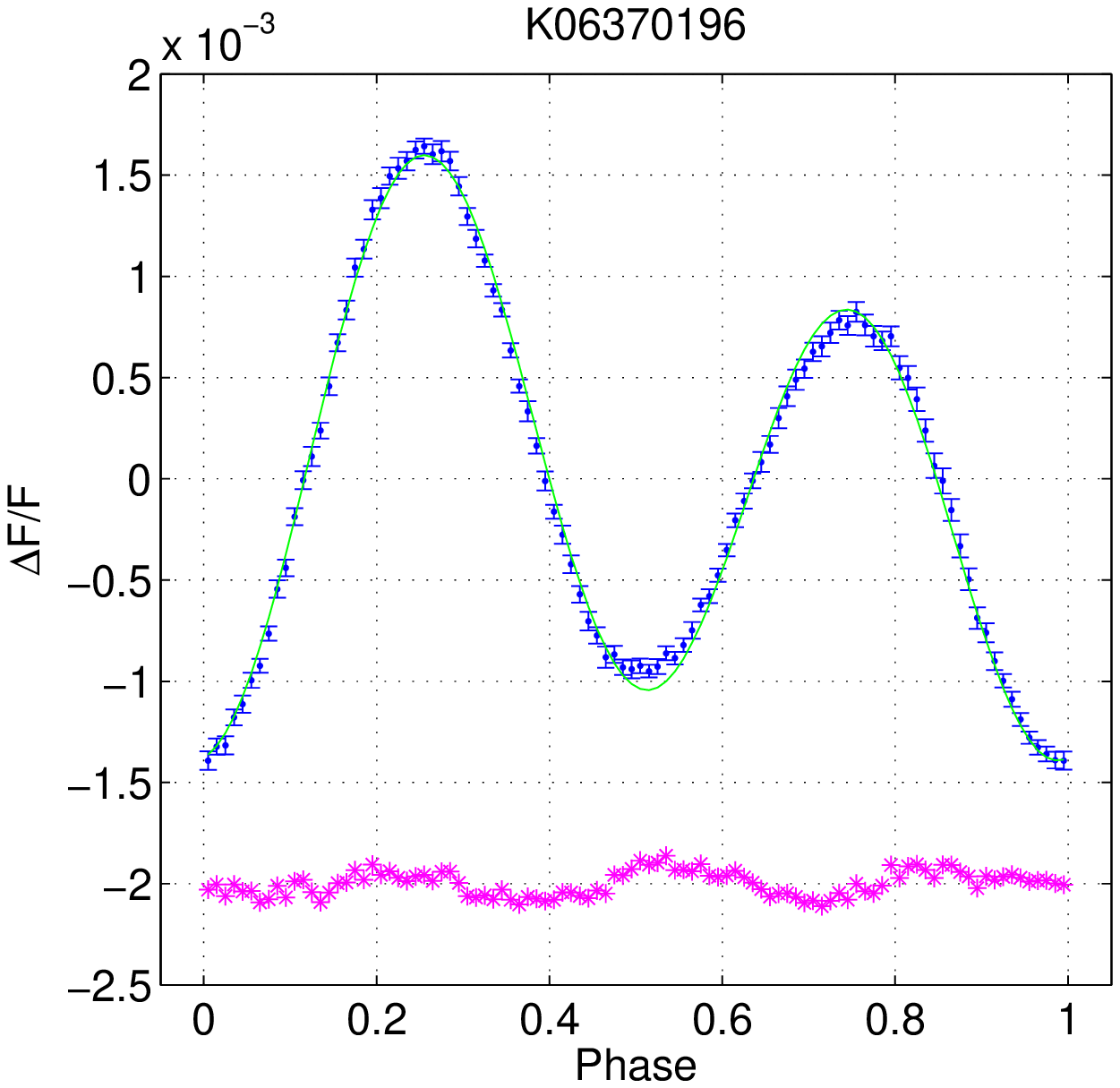}  
}

\caption{
Folded cleaned light curves, binned into 100 bins, of the seven candidates.  Phase zero is when the presumed companion is closest to the observer, while phase $0.5$ is when the primary is closest to the observer. The errors of each bin represent $1\sigma$ estimate of the bin average value. The line presents the BEER model. The model residuals are plotted at the bottom of each figure.  
}
\end{figure*}
%--------------------------------------------------------------------------------

\clearpage

%------------------------------------------
\section{Radial velocity observations}
%======================
%
 The RV observations were performed between 25 September 2010 and 15 June 2011 with the Tillinghast Reflector Echelle Spectrograph (TRES; \citet{furesz08}) mounted on the 1.5-m Tillinghast Reflector at the Fred Lawrence Whipple Observatory operated by the Smithsonian Astrophysical Observatory (SAO) on Mount Hopkins in Southern Arizona,  
using the medium fiber at a spectral resolution of $44{,}000$, covering a spectral range from 385 to 910 nm.
Exposures of a Thorium-Argon
hollow-cathode lamp immediately before and after each exposure 
were used for wavelength calibration.  The spectra were
extracted and rectified to intensity vs.\ wavelength using standard
procedures developed by Lars Buchhave \citep{buchhave10}.
 
To derive precise relative radial velocities, we performed a
cross-correlation between each observed spectrum and a template
spectrum constructed by shifting and co-adding all the observed
spectra. In addition to the template
constructed by shifting and co-adding all the observed spectra, we also tried using
the strongest individual exposure of each object as the observed template.
The two approaches gave essentially indistinguishable results, with
slightly better residuals from the orbital fits for the shifted and
co-added template.  We also derived absolute velocities using the library
of synthetic templates and found the same orbits, although with somewhat
larger residuals.

We did not include spectral orders that were significantly
contaminated by telluric lines from the Earth's atmosphere, nor did we
include the bluest orders with the lowest signal-to-noise ratio and a
few red orders with known problems.  The error of each relative
velocity was estimated using the standard deviation of the velocities
from the 21 individual orders, but the velocities themselves were
derived by first co-adding the correlation functions from the 21
orders, to get a natural weighting of the contribution from each
order. 

Using the shifted and co-added template can distort the cross-correlation peak
 because the noise in each spectrum correlates with the same noise that is
still present in the averaged template, and therefore can lead to underestimated uncertainties of the velocities. To correct this effect we later inflated the uncertainties of the orbital elements (see $\chi^2_{red}$ discussion below).

%As our template construction method may result in underestimation of the velocities uncertainties, we later inflated the orbital elements uncertainties (see $\chi^2_{red}$ discussion below).}

We used a library of synthetic spectra, calculated by John Laird for a grid of Kurucz
model atmospheres, using a line list developed by Jon Morse \citep{carney87,latham02}, to estimate values for
the effective temperature, surface gravity, metallicity, and
rotational velocity of the seven primaries. This was  done
by cross-correlating each coadded observed template spectrum against a
grid of synthetic templates surrounding the one that gave the best
correlation.  Our library of synthetic spectra has a spacing of 250 K
in effective temperature, $T_{\rm eff}$; 0.5 in log surface gravity,
$\log g$; 0.5 in the log of the metallicity compared to the sun,
[m/H]; and has a progressive spacing for rotational velocity, $v_{\rm
rot}$.  Because the grid is coarse, we used the correlation peak
heights to interpolate between grid points, arriving at a more precise
classification.  Three TRES spectral orders overlap with the synthetic
spectra, so we performed this cross-correlation and interpolation in
each order.  The mean values, weighted by the cross-correlation peak
height in each order, and RMS errors are reported in Table~2.  Note
that because of the degeneracies between $T_{\rm eff}$, $\log g$, and
[m/H] in the stellar spectra, correlated systematic errors may
dominate.  For this reason, and based on our experience in other surveys we have inflated the errors by adding 100
K in $T_{\rm eff}$ and 0.1 dex in $\log g$ and [m/H] in quadrature to
the formal order-to-order RMS errors.

% Table 2
%-------------------
%
\begin{deluxetable}{lrrrrrrr}
\tabletypesize{\scriptsize}
%\begin{table}
%\rotate
\tablecaption{ Spectra derived stellar properties of the seven binaries}

\tablewidth{0pt}
\tablehead{
& \colhead{K10848064}  & \colhead{K08016222} &  \colhead{K09512641} & \colhead{K07254760} & \colhead{K05263749} & \colhead{K04577324} & \colhead{K06370196}
}

\startdata

$T_{\rm eff}$ [K] & $6209 \pm 131$  & $5919 \pm 128$ & $6348 \pm 214$ & $6377 \pm 133$ & $6328 \pm 119$ & $6515 \pm 145$ & $6213 \pm 168$ \\
$\log g$ [dex] & $3.68 \pm 0.16$ & $4.29 \pm 0.14$ & $4.04 \pm 0.23$ & $4.04 \pm 0.23$ & $3.54 \pm 0.14$ & $3.71 \pm 0.14$ & $3.91 \pm 0.13$ \\
{[m/H]} [dex] & $-0.24 \pm 0.11$ & $-0.14 \pm 0.11$  & $-0.37 \pm 0.15$ & $-0.04 \pm 0.11$ & $-0.21 \pm 0.10$ & $-0.08 \pm 0.11$ & $-0.26 \pm 0.16$    \\
$v \sin i$ [km s$^{-1}$] & $14.26 \pm 0.45$ & $3.37 \pm 0.24$ & $9.47 \pm 0.24$ & $14.96 \pm 0.52$ & $21.50 \pm 0.16$ & $28.90 \pm 0.36$ & $20.93 \pm 0.45$ \\
\tableline

\enddata
\end{deluxetable}

The relative velocities were adjusted by a constant offset to a system
of absolute velocities using observations of the nearby IAU Radial
Velocity Standard Star HD 182488, whose absolute velocity was assumed
to be $-21.508$ km s$^{-1}$.  This adjustment utilized our library of
synthetic templates, from which we picked
the synthetic template that gave the best match to the observations of
each star in the spectral order centered on the Mg b feature near 518
nm.  This approach should avoid the problem of possible template
mis-match between the various target stars and HD 182488.  The
uncertainty in the zero point of our absolute velocities is probably
limited by the uncertainty in the absolute velocity of HD 182488, which
could be as large as 100 m s$^{-1}$. Table~3 lists the radial-velocity measurements and their uncertainties.

%---------------
% Table 3
%---------------

\begin{table}[!h]
\caption{ Radial velocities of the seven binaries}
\tiny \begin{tabular}{lrr|lrr}
\\
\hline
  Time [HJD$-2455000$]   & RV [m s$^{-1}$] & $\sigma$ [m s$^{-1}$]  &  Time [HJD$-2455000$]   & RV [m s$^{-1}$] & $\sigma$ [m s$^{-1}$] \\
\hline

K10848064:           &            &      &  K07254760: & & \\
\cline{1-6}
%=============
464.710256 &  -8835  & 119 & 694.814729 &  37083  &  39\\
469.624378 & -17974  &  76 & 695.834208 & -11894  &  46 \\
488.634305 & -16743  &  77  & 697.807122 &  15727  &  61\\
489.574441 &  -6552  & 123& 699.779976 &  46148  &  83 \\
490.617670 & -17742  &  83 & 702.796583 &  36412  &  47\\
498.599389 & -23508  &  57 & 703.815335 & -11560  &  54 \\
513.639601 &  -9108  & 170 & 704.799619 &  39179  &  51\\
692.888788 & -13130  & 124 & 705.802918 &  14063  &  89\\
\cline{4-6}
722.832949 & -14423  &  43  & K05263749: & & \\
\cline{4-6}
723.863233 &  -7328  &  56  &694.829973 &  20947  &  69\\ 
724.857438 & -21205 &  28  & 696.825053 &  13787  & 141\\
%& & & & & \\
\cline{1-3}
K08016222: & &                         & 697.826976 &  45095  &  68\\                          
\cline{1-3}
%=============
465.787007 & -31819  &  63 & 699.974074 & -12159  &  61\\
466.613973 & -37012  &  39 & 701.980887 &  34841  &  87\\
467.723417 & -33482  &  96 & 705.958593 &  23622  &  73 \\
469.613641 & -18312  &  42  & 722.851562 &  10492  &  63  \\
485.603398 & -23367  &  37 & 723.875349 &  45267  &  46 \\
490.579108 & -29588  &  63 & 724.846275 &  10122  &  75 \\
\cline{4-6}
498.591317 & -22720  &  47 &  K04577324: & &   \\
\cline{4-6}
722.843084 & -21815  &  86 & 695.982214 & -22723  & 166    \\
\cline{1-3}
K09512641: & &                         &  697.845760 &   8912  &  99 \\                    
\cline{1-3}
%===========
658.935318 &   6324  &  22 & 703.971684 &  38803  &  87\\
669.855208 &  33915  &  29 & 722.871612 &  46652  &  99 \\
693.803705 &  34608  &  67 & 723.895443 & -22140  & 103 \\
694.822698 &  16733  &  54 & 724.869757 &  34694  & 124\\
696.815346 &  17503  &  50 &  726.867639 &   5528  &  96  \\
697.816773 &  34761  &  48 &  727.847462 &  33003  &  91  \\
\cline{4-6}
698.934501 &  27404  &  39 &  K06370196: & & \\
\cline{4-6}
701.814582 &  24838  &  40 & 694.852291 &  20398  &  58\\
702.805990 &  36143  &  39 & 697.835776 & -13709  & 144 \\
703.826109 &  22562  &  59 &  702.968624 &  22603  &  41 \\
704.813460 &   6060  &  65 &  722.861284 & -35084  & 114\\
705.813161 &  11855  &  43 &  723.884266 &  17354  &  76\\
 & &                                           &  724.826295 &   8740  &  72  \\
 & &                                           &  726.857187 & -47080  &  95 \\
 & &                                           &  727.797595 &   4696  & 132 \\
\hline
\end{tabular}
\label{table_RV3}
\end{table}

For all seven candidates discussed here the first RV measurements showed clear variability. We therefore obtained enough RVs to allow orbital
solutions completely independent of the BEER analysis. 
To determine the orbital elements of each target, {\it independent of
the BEER results}, we ran a Markov chain Monte Carlo (MCMC) analysis
of the radial velocities.  We adopted values for the epoch ($T$),
period ($P$), systemic velocity ($\gamma$), orbital semi-amplitude
($K$), eccentricity ($e$), and argument of periapse ($\omega$)
corresponding to the median values of the posterior distributions.
The errors listed in the tables are those corresponding to the
$16^{th}$ and $84^{th}$ percentiles of the posterior distributions.
The reported error on $\gamma$, however, includes contributions both
from the formal error from the MCMC posterior and from the uncertainty
in the TRES absolute zero point offset.

When the orbit is circular, the epoch reported is $T_{\rm max}$, the
time of maximum velocity, and when the orbit is eccentric, we report
$T_{\rm peri}$, the time of periastron passage.  In six of seven
cases, either the orbital phase coverage was not sufficient to
adequately constrain the eccentricity or $e$ was statistically
indistinguishable from zero.  In these cases, we fixed $e=0$ and reran
the MCMC chains, adopting $T_{\rm max}$, $P$, $\gamma$, and $K$ from
this solution.  In one case, K08016222, the orbital phase coverage is good and $e$ is significantly non-zero.

%---------------------------
% Figure 4%
%---------------------------
\begin{figure*} 
\centering
\resizebox{16cm}{6cm}
{
 \includegraphics{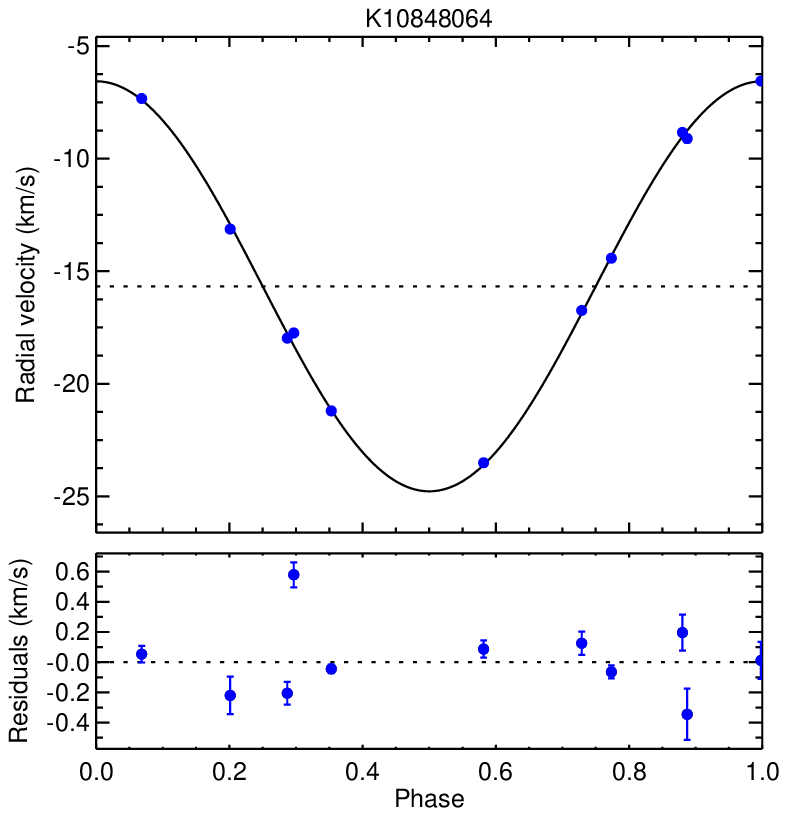}  
 \includegraphics{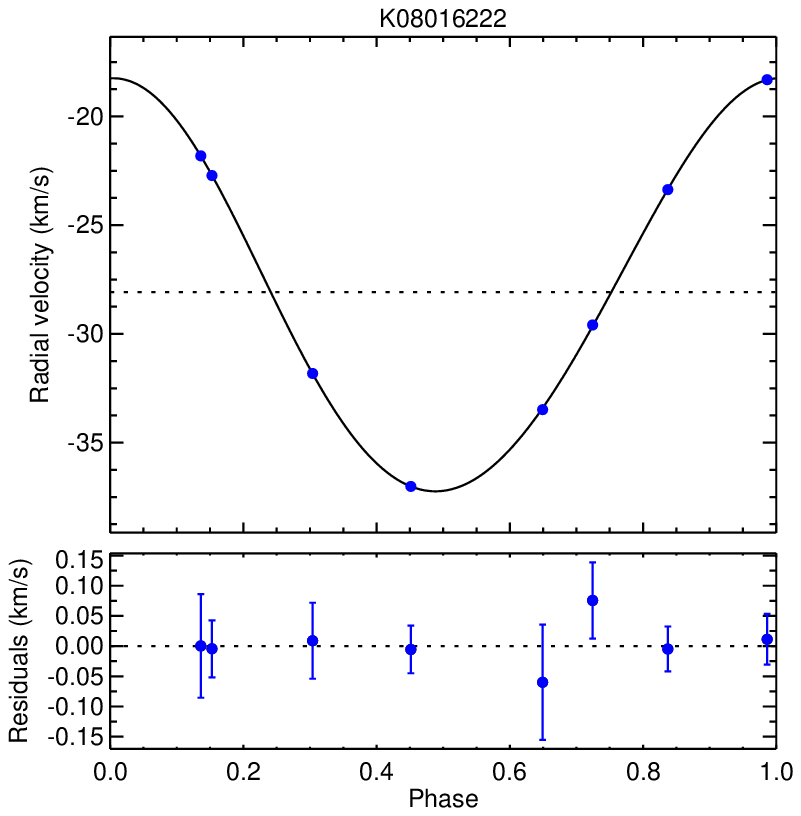}
 \includegraphics{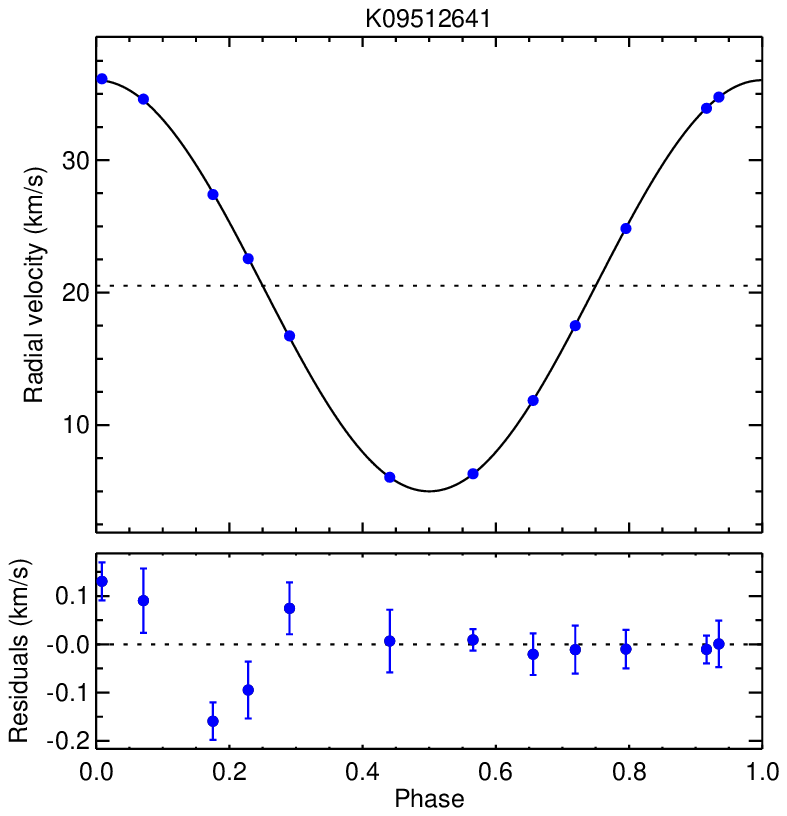}
}

\centering
\resizebox{16cm}{6cm}
{
 \includegraphics{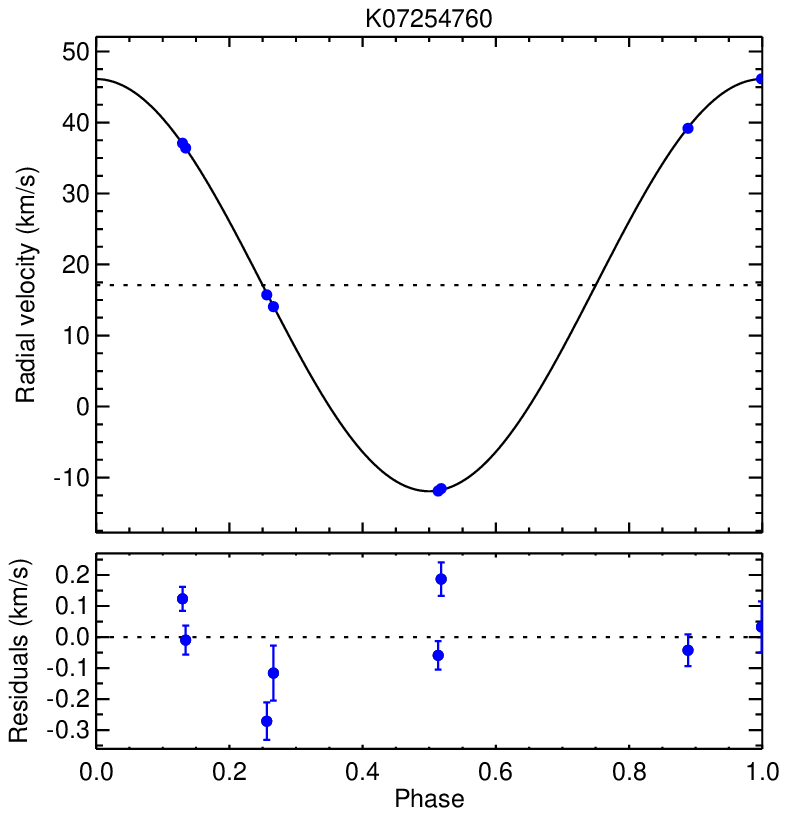}  
 \includegraphics{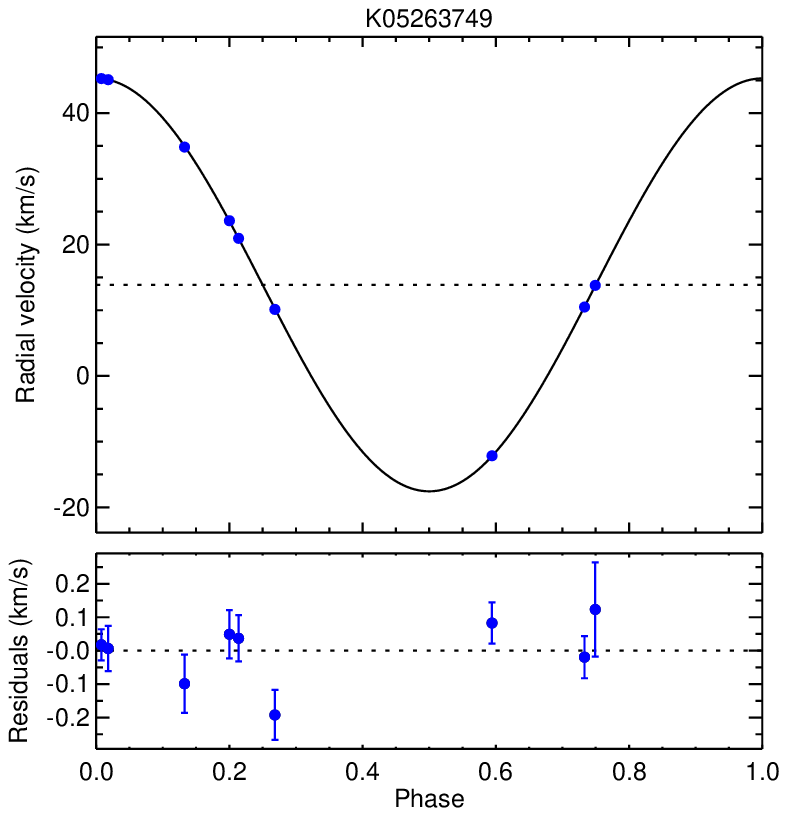}
 \includegraphics{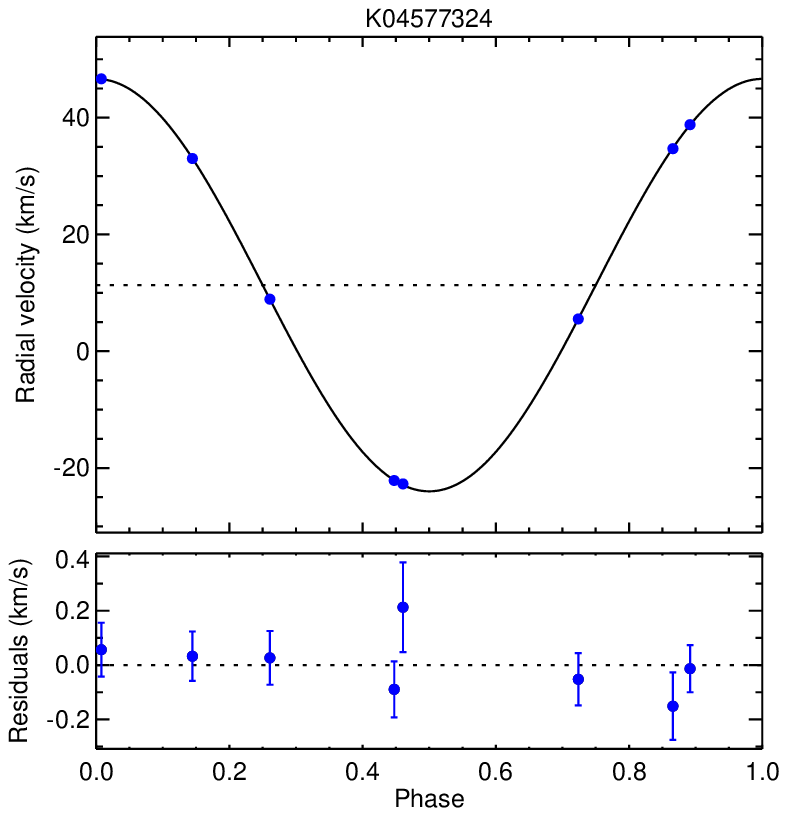}
}

\centering
\resizebox{5.33cm}{6cm}
{
 \includegraphics{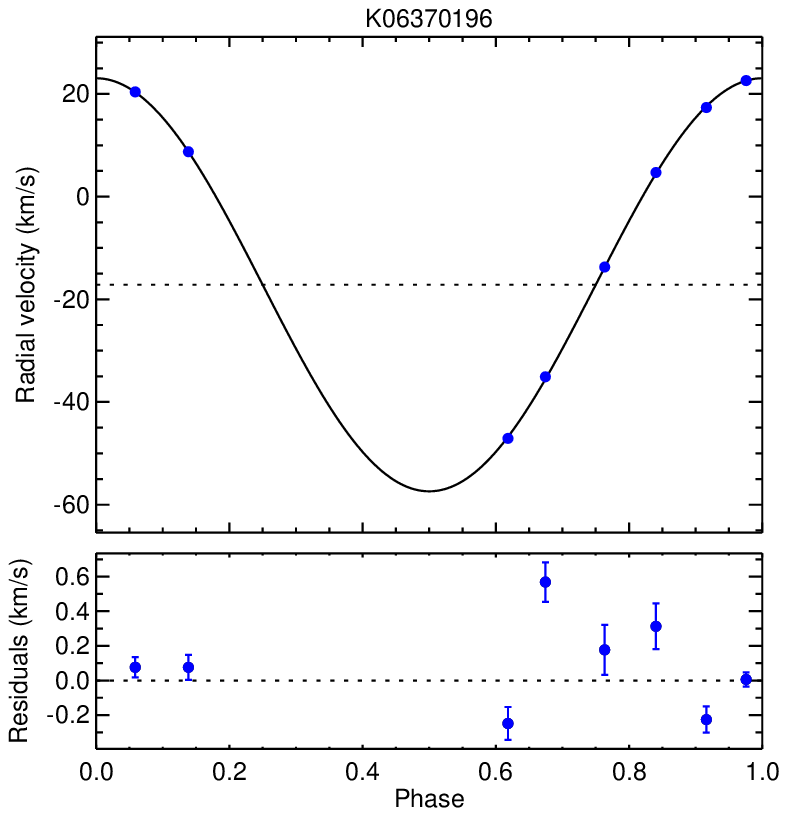}  
}

\caption{
The RV measurements of the seven stars, folded at the derived orbital period. The line presents the orbital RV model. The residuals are plotted at the bottom of each figure.  Note the different scale of the upper and lower panel of each star. The error bars are too small to be seen in the upper panels. 
}
\end{figure*}
%--------------------------------------------------------------------------------

Figure~4 shows the RV follow-up measurements for each of the seven binaries, folded with the period found, and  Table~4 lists the derived orbital
elements. The table also lists $\chi^2_{red}$, the reduced $\chi^2$ of the model, and the time span of the observations. For two binaries the derived $\chi^2_{red}$ value is close to unity, as expected, but for the others its value is relatively large. This could indicate either that for those binaries our radial-velocity uncertainties are underestimated, or that our radial-velocity model is too simple, due to  some stellar noise, for example.
In order to get more realistic uncertainties for the model elements, we inflated the parameter uncertainties of each target by its $\sqrt{\chi^2_{red}}$, which is equivalent to inflating the RV errors of that star by the same factor. The resulting orbital model elements uncertainties are listed in Table~4.

%
% Table 4: Orbital model elements of the seven binaries
%-----------------------------------------------------------------
%
\begin{deluxetable}{lrrrrrrr}
\tabletypesize{\scriptsize}
%\begin{table}
%\rotate
\tablecaption{ Orbital model elements of the seven binaries}

\tablewidth{0pt}
\tablehead{
& \colhead{K10848064} & \colhead{K08016222} &  \colhead{K09512641} & \colhead{K07254760} & \colhead{K05263749} & \colhead{K04577324} & \colhead{K06370196}
}
\startdata
 $T_{\rm max}-2455000$  [HJD]&  $465.1289$  & $464.6288^{a}$  & $642.3690$  & $702.4395$  & $466.7071$  & $466.7005$  & $698.8374$ \\
  & $\pm 0.0060$ & $\pm 0.0029$ & $\pm 0.0052$ & $\pm 0.0126$ & $\pm 0.0124$ & $\pm 0.0079$ & $\pm 0.0035$  \\
\\
$P$ [days] & $3.49318 $ & $5.60864$  &$4.64588 $ &$2.65642 $ &$3.72665 $ &$2.328663 $ &$4.23371 $\\
 & $\pm 0.00099$ &  $\pm0.00017$ & $\pm 0.00044$ & $\pm 0.00068$ & $\pm 0.00019$ & $\pm 0.000070$ & $\pm 0.00067$ \\
\\
 $\gamma$  [km s$^{-1}$]& $-15.670 $ &$-28.078 $ &$20.518 $ &$17.092 $ &$13.862$ &$11.325 $ & $-17.168 $\\
 & $\pm 0.219$  & $\pm 0.048$ & $\pm 0.110$ & $\pm 0.177$ &  $\pm 0.060$ & $\pm 0.119$ & $\pm 0.454$  \\
\\
$K$  [km s$^{-1}$] & $9.107 $ &$9.495 $ & $15.519 $ &$29.024 $ &$31.428 $ &$35.316 $ &$40.222 $ \\
&   $\pm 0.073$ & $\pm 0.018$ & $\pm 0.023$ & $\pm 0.061$ & $\pm 0.040$ & $\pm 0.043$ & $\pm 0.131$ \\
\\
$e$ & $0$ (fixed) & $0.0439 \pm 0.0022$  & $0$ (fixed) & $0$ (fixed) & $0$ (fixed) & $0$ (fixed) & $0$ (fixed)\\
$\omega$    [deg] & & $36.2 \pm 2.6$            & & & & & \\
 $\chi^2_{red}$ & $11.0$ & $1.0$  & $4.3$ & $11.6$ & $2.6$ & $1.2$ & $12.6$\\
span  [days]   &  $260.1$ &    $257.1$ &  $46.9$ &  $11.0$ &  $30.0$ &  $31.9$ & $32.9$ \\

\tableline
\enddata
\tablenotetext{a}{ for K08016222 the $T_{\rm max}$ value is the time of periastron passage }
\end{deluxetable}
%---------------------------------------------------------------------------------------------

%============%
\section{Results}        %
%============%

Table~5 lists for each of the seven newly discovered binaries the period derived from the photometry, the calculated $\alpha_{beam}$, and the expected RV semi-amplitude, $K_{beam}$, derived from $\alpha_{beam}$ and the photometric beaming amplitude.
% the amplitude of the beaming effect through the $\alpha_{beam}$ factor \citep{faigler11}, for which we assumed value of unity, that is valid under the bolometric flux assumption}.
The $\alpha_{beam}$ factor includes one component that originates from the fact that the stellar spectrum is Doppler shifted relative to the observed band. To estimate this factor for each of the seven detected binaries we numerically shifted spectra from the library of \citet{castelli04} models that were close to the estimated temperature, metallicity and gravity of each of the seven stars. The values adopted were derived by interpolation of the $\alpha_{beam}$ values between the available models of the library.  
%To estimate $\alpha_{beam}$ we used the \cite{castelli04} atmospheric model of each of the targets to estimate the stellar spectrum, and integrated the Doppler shifted stellar spectrum over the {\it Kepler} bandpass \citep{zucker07}
 The $\alpha_{beam}$ uncertainties were estimated by calculating the interpolated $\alpha_{beam}$ values within the $T_{\rm eff}$, $\log g$ and [m/H] error ranges. 
The error on the expected $K_{beam}$ was estimated by combining the photometric beaming amplitude error and the $\alpha_{beam}$ error.
%We estimate the error on the expected amplitude to be on the order of 25\%. 
The table then reports the number of RV measurements, their derived RV period and semi-amplitude, and the minimal secondary mass, up to $\sin i$.  For all cases we {\it independently} derived the period of the RV modulation, and found it to be consistent with the photometric period, indicating that the orbital period was reliably derived by the BEER algorithm, solely from the photometric data.

%---------
%Table 5
%-----------

\begin{deluxetable}{lrrrrrrr}
\tabletypesize{\scriptsize}
%\begin{table}
%\rotate
\tablecaption{
Derived photometric RV period and semi-amplitude together with RV observations period and semi-amplitude for each of the seven binaries}
\tablewidth{0pt}
\tablehead{
% { \small \begin{tabular}{lccccccc}
& \colhead{K10848064}  & \colhead{K08016222} &  \colhead{K09512641} & \colhead{K07254760} & \colhead{K05263749} & \colhead{K04577324} & \colhead{K06370196}
}
% & K6222  & K8064 &  K2641 & K4760 & K3749\\
\startdata

%Stellar mass       & $1.1$ & $1.2$ & $1.2$ & $1.2$ & $1.3$ & $1.2$ & $1.3$         & $ M_{\odot}$  \\
%Stellar radius       & $1.3$ & $1.5$ & $1.7$ & $1.5$ & $1.9$ & $1.3$ & $2.1$         &  $ R_{\odot}$\\
%{\it Kepler} magnitude & $11.6$ & $12.1$ & $11.7$ & $12.0$ & $11.5$ & $12.0$ & $12.0$    & mag  \\
%\tableline
Photometry results:\\

Period [days]   & $3.49 \pm 0.01$  & $5.60 \pm 0.02$ & $4.65 \pm 0.02$ & $2.66 \pm 0.01$ & $3.73 \pm 0.01$ & $2.33 \pm 0.01$ & $4.23 \pm 0.01$  \\ 

%$\alpha_{\mathrm\scriptscriptstyle beam}$ & 0.44 & 0.50 & 0.43 & 0.44 & 0.44 & 0.39 & 0.43  \\
%Expected $K_{\mathrm\scriptscriptstyle beam}$ [km s$^{-1}$] & $18$ & $10$ & $22$ & $56$ & $56$ & $80$ & $58$      \\   
%Expected $K_{\mathrm\scriptscriptstyle beam}$ [km s$^{-1}$] & $9$ & $5$ & $10$ & $27$ & $26$ & $33$ & $27$      \\   
%$\alpha_{beam}$ & $0.944 \pm 0.025$ & $1.012 \pm 0.023$ & $0.912 \pm 0.036$ & $0.921 \pm 0.024$ & $0.921 \pm 0.022$ & $0.887 \pm 0.025$ & $0.936 \pm 0.03$ \\ 
$\alpha_{beam}$ & $0.944 \pm 0.025$ & $1.012 \pm 0.023$ & $0.912 \pm 0.036$ & $0.921 \pm 0.024$ & $0.921 \pm 0.022$ & $0.887 \pm 0.025$ & $0.936 \pm 0.030$ \\ 
$K_{\mathrm\scriptscriptstyle beam}$ [km s$^{-1}$] & $9.37 \pm 0.34$ & $7.19 \pm 0.22$ & $15.21 \pm 0.72$ & $28.97 \pm 0.89$ & $29.14 \pm 0.81$ & $36.86 \pm 1.10$ & $30.61 \pm 1.12$      \\   
%Expected $K_{\mathrm\scriptscriptstyle beam}$ [km s$^{-1}$] & $8\pm 2 $ & $5 \pm 1.2$ & $10 \pm 2.5$ & $25 \pm 6.3$ & $24 \pm 6$ & $31 \pm 7.8 $ & $25 \pm 6.3$      \\   
\\
 \tableline
\\
RV results:\\
$N_{obs}$  & $11$  & $8$ & $12$ & $8$ & $9$ & $8$ & $8$ \\
\\
Period  [days]  & $3.49318$   & $5.60864$& $4.64588$  & $2.65642$  & $3.72665$  & $2.328663$  & $4.23371$  \\ 

  & $\pm 0.00099$ &  $\pm0.00017$ & $\pm 0.00044$ & $\pm 0.00068$ & $\pm 0.00019$ & $\pm 0.000070$ & $\pm 0.00067$ \\
\\
 $K_{\mathrm\scriptscriptstyle RV}$ [km s$^{-1}$] & $9.107 $  & $9.495$  & $15.519 $ & $29.024 $ & $31.428 $ & $35.316 $ & $40.223$    \\

&   $\pm 0.073$ & $\pm 0.018$ & $\pm 0.023$ & $\pm 0.061$ & $\pm 0.040$ & $\pm 0.043$ & $\pm 0.131$ \\
\\
Minimum secondary  & $76 \pm 5$  & $90 \pm 6$ & $147 \pm 10$ & $222 \pm 15$ & $279 \pm 19$ & $253 \pm 17$ &  $376 \pm 25$ \\
mass [$M_{\mathrm Jup}$]\\
%\tableline
\enddata
\end{deluxetable}
%--------------------------------------------------------------

%-----------------------------
In six of the binaries the eccentricity was too small to be derived significantly, so we assumed circular orbits.
Because these are short-period stellar binaries, the expected circularization timescale is short, 
so finding in most cases that $e=0$ is consistent with our expectations.  For K08016222 we find $e=0.0439\pm0.0022$. Interestingly, this is the binary with the longest period, so its lifetime might have been too short to achieve circularization \citep{mathieu88}. 
%---------------------------

Out of the seven binaries, the measured RV amplitudes of five cases were consistent with those predicted by the photometric analysis. For the other two stars, K08016222 and K06370196, the predicted amplitudes were $24\%$ smaller than the observed ones. This could be due to underestimation of the photometric amplitude. Another possible explanation may be an inaccurate translation of the photometric amplitude to the expected RV amplitude, which depends on the assumed stellar spectral type.  We need more confirmed binaries to understand this effect.

\section{Discussion}
The RV observations presented here demonstrate the ability of the BEER algorithm to discover short-period binaries with minimum secondaries mass in the range of $0.07$--$0.4\, M_{\odot}$ in the publicly available {\it Kepler} data. 

The original goal of the Kepler and CoRoT missions was to search for transiting planets. Such projects are limited to planets with orbital inclinations close to $90^{\circ}$. The serendipitous discoveries of eclipsing binaries in the {\it Kepler} photometry \citep{prsa10} are suffering from the same limitation. The BEER algorithm, on the other hand, is searching for {\it non-transiting} companions, and therefore can detect many more systems with much lower inclination angles. Searching with BEER is effectively equivalent to performing  an RV survey that is not limited to nearly face-on inclinations. Applying the BEER algorithm to the hundreds of thousands of already available light curves of {\it Kepler} and CoRoT is like performing an RV survey of a huge sample that is composed of these stars.  

Therefore, we expect BEER to discover many hundreds of new binaries with short periods. Furthermore, whereas in RV studies the actual mass of the companion depends on the unknown inclination angle,
detecting {\it both} the ellipsoidal and the beaming effects will enable BEER  to derive, or at least estimate, the mass of the small companion in certain cases. As pointed out by \citet{faigler11}, this can become possible because the two effects have different dependencies on the orbital inclination, and therefore the derived ratio of the amplitudes of the two effects can, {\it in principle}, remove the degeneracy between the secondary mass and the inclination.   

Obviously, at this stage of the BEER search, detecting a candidate is not enough --- the candidates have to be confirmed by follow-up RV observations. However, when we accumulate enough observations we will be able to estimate the false alarm probability, which might be a function of the amplitude of the photometric modulation and the stellar mass, radius and temperature. Therefore, we will be able to derive the statistical features of the short-period binaries without confirming each detection with RV observations.   

The seven cases presented here were based on the Kepler Q0--Q2 data. 
\citet{faigler11} suggested that once the full {\it Kepler} dataset is available, we should be able to detect brown-dwarf secondaries  and even massive planets. Moreover, the other stellar modulations that contribute now to the false alarm frequency are not expected to be so stable on time scales of years, whereas the three BEER effects are strictly periodic and stable. Therefore, we expect the false alarm frequency to decrease when we have access to longer data sets. The unprecedentedly large sample size and data quality, together with a knowledge of the false alarm probability, could serve as a tool to study accurately the frequency of low-mass secondaries in short-period binaries on the high- and low-mass ends of the brown-dwarf desert \citep{raghavan10, udry10, sahlmann10}. 

\acknowledgments
We are indebted to Shay Zucker and Ehud Nakar for helpful discussions. We thank the anonymous referee for highly illuminating comments and suggestions.

We feel deeply indebted to the team of the Kepler mission, that enabled us to search and analyze their unprecedentedly accurate photometric data.

All the photometric data presented in this paper were obtained from the 
Multimission Archive at the Space Telescope Science Institute (MAST). 
STScI is operated by the Association of Universities for Research in 
Astronomy, Inc., under NASA contract NAS5-26555. Support for MAST for 
non-HST data is provided by the NASA Office of Space Science via grant 
NNX09AF08G and by other grants and contracts.

We thank the Kepler mission for partial support of the spectroscopic
observations under NASA Cooperative Agreement NNX11AB99A with the
Smithsonian Astrophysical Observatory, DWL PI.

This research was supported by the ISRAEL SCIENCE FOUNDATION (grant No.
655/07).

%% To help institutions obtain information on the effectiveness of their
%% telescopes, the AAS Journals has created a group of keywords for telescope
%% facilities. A common set of keywords will make these types of searches
%% significantly easier and more accurate. In addition, they will also be
%% useful in linking papers together which utilize the same telescopes
%% within the framework of the National Virtual Observatory.
%% See the AASTeX Web site at http://www.journals.uchicago.edu/AAS/AASTeX
%% for information on obtaining the facility keywords.

%% After the acknowledgments section, use the following syntax and the
%% \facility{} macro to list the keywords of facilities used in the research
%% for the paper.  Each keyword will be checked against the master list during
%% copy editing.  Individual instruments or configurations can be provided 
%% in parentheses, after the keyword, but they will not be verified.

{\it Facilities:} \facility{FLWO:1.5m (TRES)}

%% The reference list follows the main body and any appendices.
%% Use LaTeX's thebibliography environment to mark up your reference list.
%% Note \begin{thebibliography} is followed by an empty set of
%% curly braces.  If you forget this, LaTeX will generate the error
%% "Perhaps a missing \item?".
%%
%% thebibliography produces citations in the text using \bibitem-\cite
%% cross-referencing. Each reference is preceded by a
%% \bibitem command that defines in curly braces the KEY that corresponds
%% to the KEY in the \cite commands (see the first section above).
%% Make sure that you provide a unique KEY for every \bibitem or else the
%% paper will not LaTeX. The square brackets should contain
%% the citation text that LaTeX will insert in
%% place of the \cite commands.

%% We have used macros to produce journal name abbreviations.
%% AASTeX provides a number of these for the more frequently-cited journals.
%% See the Author Guide for a list of them.

%% Note that the style of the \bibitem labels (in []) is slightly
%% different from previous examples.  The natbib system solves a host
%% of citation expression problems, but it is necessary to clearly
%% delimit the year from the author name used in the citation.
%% See the natbib documentation for more details and options.

\end{document}